\documentclass[aps,prb,reprint,nofootinbib,longbibliography,superscriptaddress]{revtex4-2}
\usepackage{graphicx,graphics,color,epsfig}
\usepackage{bm}
\usepackage{dcolumn}
\usepackage{amsmath}
\usepackage{amssymb}
\usepackage{graphicx}
\usepackage{epstopdf}
\usepackage[colorlinks=true,  
    linkcolor=blue,          
    citecolor=blue,          
    urlcolor=blue,           
    anchorcolor=blue]{hyperref}
\usepackage{multirow}  
\usepackage{lineno} 
\usepackage[normalem]{ulem} 

\newcommand{\be}{\begin{eqnarray}}
\newcommand{\ee}{\end{eqnarray}}

\newcommand{\ba}{\begin{align}}
\newcommand{\ea}{\end{align}}

\makeatletter

\newcommand{\Rmnum}[1]{\expandafter\@slowromancap\romannumeral #1@}
\makeatother

\begin{document}

\title{Exploring Dynamical Quantum Phase Transitions via the Dynamical Phase \\ Accumulated at an Equilibrium Critical Point}


\affiliation{School of Sciences, Southwest Petroleum University,
Nanchong 637001, China}

\affiliation{College of Science, Sichuan Agricultural University,
Ya'an 625014, China}

\affiliation{College of Physics, Sichuan University,
Chengdu 610064, China}


\author{Han-Chuan Kou}
\affiliation{School of Sciences, Southwest Petroleum University,
Nanchong 637001, China}

\author{Peng Li}
\affiliation{College of Physics, Sichuan University,
Chengdu 610064, China}

\author{Zhi-Han Zhang}
\affiliation{College of Physics, Sichuan University,
Chengdu 610064, China}

\author{Zhen-Yu Zheng}
\email{zhengzhenyu@sicau.edu.cn}
\affiliation{College of Science, Sichuan Agricultural University,
Ya'an 625014, China}

\date{\today}

\begin{abstract}
 We investigate dynamical quantum phase transitions (DQPTs) generated by a two-step linear quench with a finite waiting time at the equilibrium critical point. The waiting stage leads to the accumulation of a dynamical phase that depends on the waiting time and directly reflects the critical spectral structure. In the transverse Ising chain, this accumulated phase modifies the critical structure of DQPTs from a single-critical-mode (SCM) regime to a multiple-critical-mode (MCM) regime. We analytically and numerically determine the boundary between the two regimes. Along this boundary, the waiting time approaches a constant in the sudden-quench limit, whereas in the slow-quench limit it follows Kibble--Zurek (KZ) scaling with the quench time. In the SCM regime, the rate function exhibits a conventional linear singularity, $r(t)-r(t^{*})\propto |t-t^{*}|^\alpha$, with the critical exponent $\alpha=1$ near the critical time. At the SCM--MCM boundary, this exponent changes to $\alpha=1/2$, leading to a divergent first derivative of the rate function. Additional discrete parameter lines with $\alpha=1/2$ also appear within the MCM regime in the parameter plane spanned by the waiting and quench times. Our findings establish a direct connection between the accumulated dynamical phase and the critical structure of DQPTs, further supported by an extended Su--Schrieffer--Heeger (SSH) model.
\end{abstract}

\maketitle


\section{Introduction}

Dynamical quantum phase transitions (DQPTs) are characterized by nonanalytic behavior in the real-time evolution of quantum systems. The Loschmidt amplitude $\mathcal{G}=\langle\psi|e^{-iHt}|\psi\rangle$ plays a role analogous to the equilibrium partition function, and its nonanalyticities can be examined through Fisher zeros \cite{Hely_2013, Heyl_2018}. During the past decade, DQPTs have attracted extensive theoretical interest in a wide variety of systems, including integrable spin chains \cite{Dirk_2016, Dutta_2018, Heyl_2018-2, Ding_2020, Polkovnikov_2021, Cao_2023, LiFX_2025, Cao_2026}, topological systems \cite{Huang_2016, Chi_2024-apr, Chi_2024}, interacting many-body models \cite{Markus_2015, Hely_2017, Peotta_2021PRX}, and open quantum systems \cite{Link_2020}. Meanwhile, experimental observations of DQPTs have been reported in a variety of quantum platforms, spanning trapped ions \cite{Hely_2017, Bian_2026}, quantum walks \cite{XueP_2019, GuoGC_2020, XueP_2025}, ultracold fermionic gases \cite{Hely_2018}, and spinor Bose--Einstein condensates \cite{DuanLM_2020}. Additionally, superconducting qubit simulators \cite{ChenS_2019, XuK_2020, Green_2022} and digital Rydberg-atom platforms \cite{Elisa_2026} have emerged as versatile tools for simulating DQPTs and probing their dynamical and topological properties. 

The concept of DQPTs was first introduced by Heyl in the context of a sudden quench \cite{Hely_2013}. In many systems, DQPTs at a critical time $t^*$ are accompanied by quantized changes in a dynamical topological order parameter constructed from the Pancharatnam geometric phase \cite{Hely_2016_DTOP}. The Loschmidt rate function can also exhibit scaling behavior characterized by a critical exponent $\alpha$, $r(t)-r(t^*)\propto |t-t^*|^\alpha$ \cite{Hely_2015PRL, WuYT_2020}.
A complementary framework for nonequilibrium critical dynamics is the Kibble--Zurek mechanism (KZM), which describes universal scaling when a system is slowly driven across a quantum critical point \cite{Zurek_2005, Dziarmaga_2005, Polkovnikov_2005, kou_2022, Campo_2023, LiP_2023, Ding_2024, YinS_2026, Kiss_2025, kou_2025, LiP_2025, Kou_2026}. The resulting post-quench state may exhibit DQPT phenomena distinct from those following a sudden quench  \cite{Polkovnikov_2016, Jafari_2024May, Jafari_2024Aug, CaoKY_2024, ZhengZY_2025, Jafari_2025}. In the transverse Ising chain, Fisher zeros depend on the excitation probability, while the associated real-time singularities are linked to critical modes. However, the nonequilibrium evolution is characterized not only by the excitation probability but also by an accumulated dynamical phase \cite{Nori_2010, Nori_2023}. Recent studies have shown that the evolution time between two successive quenches can strongly modify the interference conditions of the Loschmidt amplitude, enabling the control of DQPTs through the accumulated dynamical phase \cite{Porta_2020, Kennes_2018, Chien_2022, Cheraghi_2023}. 

However, how a dynamical phase accumulated specifically at an equilibrium critical point restructures the critical modes and modifies the local singularity of DQPTs remains unclear. To address this question, we consider a two-step linear quench with a finite waiting stage at an equilibrium critical point. The gapless critical spectrum allows low-energy modes to accumulate dynamical phases during the waiting stage. These phases differ from those acquired during a linear quench through the critical region \cite{Dziarmaga_2010, Dziarmaga_2021, Dziarmaga_2022}. In both the transverse Ising and extended SSH models, we demonstrate that the phases accumulated during the waiting stage give rise to coherent oscillations in the excitation probability of the post-quench state. By tuning the quench time and waiting time, these oscillations give rise to additional pairs of DQPT critical modes, thereby driving the system from the SCM regime to the MCM regime. We determine the SCM--MCM boundary in both the fast- and slow-quench limits. Along this boundary, the waiting time approaches a constant in the sudden quench limit and exhibits KZ scaling in the slow quench case. Remarkably, a pair of critical modes coalesces at the boundary into a tangential critical mode, changing the local singularity of the rate function from the conventional linear cusp with $\alpha=1$ to a square-root form with $\alpha=1/2$, whose first derivative diverges.

The remainder of this paper is organized as follows. In Sec. \ref{sec-Ising}, we introduce the two-step quench protocol and analyze the excitation probability modified by the accumulated dynamical phase. In Sec. \ref{sec-dqpt}, we characterize the SCM and MCM regimes and determine their boundary. In Sec. \ref{sec-ratefunc}, we investigate the rate function and its critical behavior. In Sec. \ref{sec-ESSH}, we investigate the DQPTs in the extended SSH model. Finally, Sec. \ref{sec-conclusion} summarizes the main results and discusses their implications.

\begin{figure}[t]
  \begin{center}
		\includegraphics[width=3.3 in,angle=0]{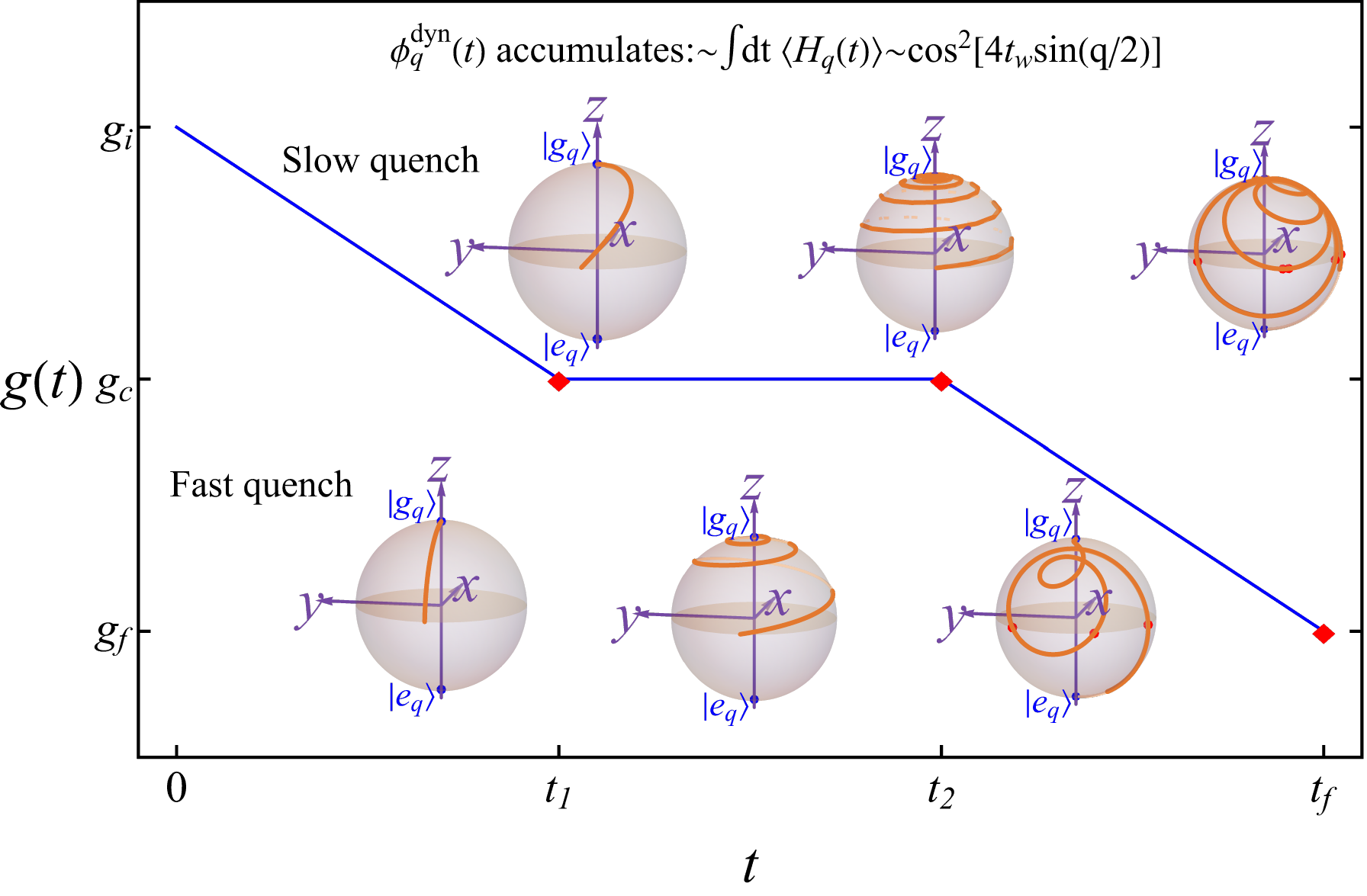}
  \end{center}
  \caption{Quench protocol defined in Eq. \eqref{eq-protocol} and the accumulation of dynamical phase during the waiting stage. 
  The upper Bloch spheres correspond to the slow quench with $\tau_Q=8$ and $t_w=16$. The lower ones correspond to the fast quench with $\tau_Q=0.01$ and $t_w=2$. In each group, the three Bloch spheres represent the quantum states at $t_1$ and $t_2$ and $t_f$. $|g_q\rangle$ and $|e_q\rangle$ are instantaneous ground and excited states of $H_q$. The dynamical phase accumulated at the critical point produces coherent oscillations in the final state. At $t_f$, the Bloch vectors cross the $x-y$ plane (i.e., $p_q=1/2$) more than once, indicating the emergence of multiple critical modes during the subsequent free evolution.}
  \label{plot-evolution}
\end{figure}

\section{Transverse Ising chain and quench protocol}\label{sec-Ising}

We start from the transverse Ising chain
\begin{align}\label{H-Ising}
H=-\sum_{j=1}^{N}\left( \sigma_{j}^{x}\sigma_{j+1}^{x}+g\sigma_{j}^{z}\right),
\end{align}
where $\sigma^{a}_j$ ($a=x,y,z$) are Pauli matrices. $N$ is the total number of lattice sites and is assumed to be even. We consider ferromagnetic coupling and impose periodic boundary conditions, $\sigma^{a}_{N+j}=\sigma^{a}_{j}$. $g$ is the strength of the transverse field. 

The system is initially prepared in the ground state corresponding to the transverse field $g_i$ at $t_i=0$, after which the field is varied to its final value $g_f=0$ at $t_f$. Here, we adopt a two-step linear quench protocol with a waiting time $t_w$ at the critical point $g_c$,
\begin{align} \label{eq-protocol}
 g(t)=\left\{
 \begin{array}{cc}
 g_i-\frac{t}{\tau_Q},
 & 0\leq t<t_1,\\
 g_c,
 & t_1 \leq t\leq t_2,\\
 g_i-\frac{t-t_w}{\tau_Q},
 &  t_2 \leq t < t_f,
 \end{array}\right.
\end{align}
where $t_1=(g_i-g_c)\tau_Q$, $t_2=t_1+t_w$, and $t_f= g_i\tau_Q+t_w$. $\tau_Q$ is the quench time, and the quench protocol is shown in Fig. \ref{plot-evolution}. In general, the nonequilibrium critical dynamics can be captured by excitation probability $p_q$ and defect density $n$, as defined in App. \ref{App-sec-KZM}. 

The dynamical phase accumulated during the waiting stage leads to coherent oscillations in the final excitation probability. Fig. \ref{plot-evolution} illustrates this effect through Bloch-sphere trajectories for representative slow and fast quenches. The resulting interference is encoded in the final excitation probability, which reads
\begin{align}
 p_{q}&=p_q^\text{half}\cos^2(\sqrt{\pi\tau_Q}q + \phi_q^\text{w}),~\tau_Q\gg 1, \label{eq-pqKZS-waiting}\\
 p_q&=p_q^\text{su}\cos^2\phi_q^\text{w},~\tau_Q\rightarrow 0,   \label{eq-nsudd-waiting}
\end{align}
where 
\begin{align}
  p_q^\text{half}=&4p_q^\text{lz}(t_1)\left[1-p_q^\text{lz}(t_1)\right],\label{eq-nKZS-half}\\
  p_q^\text{su}=&\cos^2\frac{q}{2} -\frac{1}{2g_i}\sin^2 q.  \label{eq-pqsudden}
\end{align}
$p_q^\text{lz}(t_1)$ denotes the excitation probability at $t_1$ in the long-wavelength approximation and is given by
\begin{align}\label{eq-pqlzt1}
  p_q^\text{lz}(t_1)=&\frac{1}{2}+\pi\sqrt{\frac{\tau_Q q^2}{2}}e^{-\frac{1}{2}\pi\tau_Q q^2}\nonumber\\
  &\times \Re\left[\frac{e^{-i\frac{3\pi}{4}}}{\Gamma\left(\frac{1}{2}-\frac{i\tau_Q q^2}{2}\right)\Gamma\left(1+\frac{i\tau_Q q^2}{2}\right)}\right],
\end{align}
where $\Gamma(x)$ is the gamma function. Details are provided in App. \ref{App-pq-waiting}. There are two phase factors, $\sqrt{\pi\tau_Q}q$ and $\phi_q^\text{w}$, appearing in Eqs. \eqref{eq-pqKZS-waiting} and \eqref{eq-nsudd-waiting}. They originate from distinct dynamical processes and play different roles in the nonequilibrium evolution. The former arises from the passage through the critical region in the two linear quench stages, and its expression is obtained within the long-wavelength approximation \cite{Sen_2010}. The latter is the accumulated dynamical phase during the waiting time at the critical point,
\begin{align} \label{eq-accuDP}
  \phi_q^\text{w}=-\int_{0}^{t_w}\langle g_q(g_c)|H_q(g_c)|g_q(g_c)\rangle dt=t_w \omega_q(g_c)
\end{align}
where $H_q$ is defined in Eq. \eqref{eq-hqmode}, $|g_q\rangle$ is the ground state of $H_q$, and $\omega_q(g_c)=4 \sin\frac{q}{2}$ is the spectrum of $H_q$ at the critical point $g_c=1$. Whereas the KZM characterizes the excitation probability and defect density generated by crossing the critical point, the accumulated phase $\phi_q^\text{w}$ during the waiting stage retains information about the critical dispersion. Its momentum dependence produces coherent interference in $p_q$, which in turn restructures the DQPT critical modes. 

We next discuss several features of these coherent oscillations in the excitation probability and defect density.

\begin{figure}[t]
  \begin{center}
		\includegraphics[width=3.2 in,angle=0]{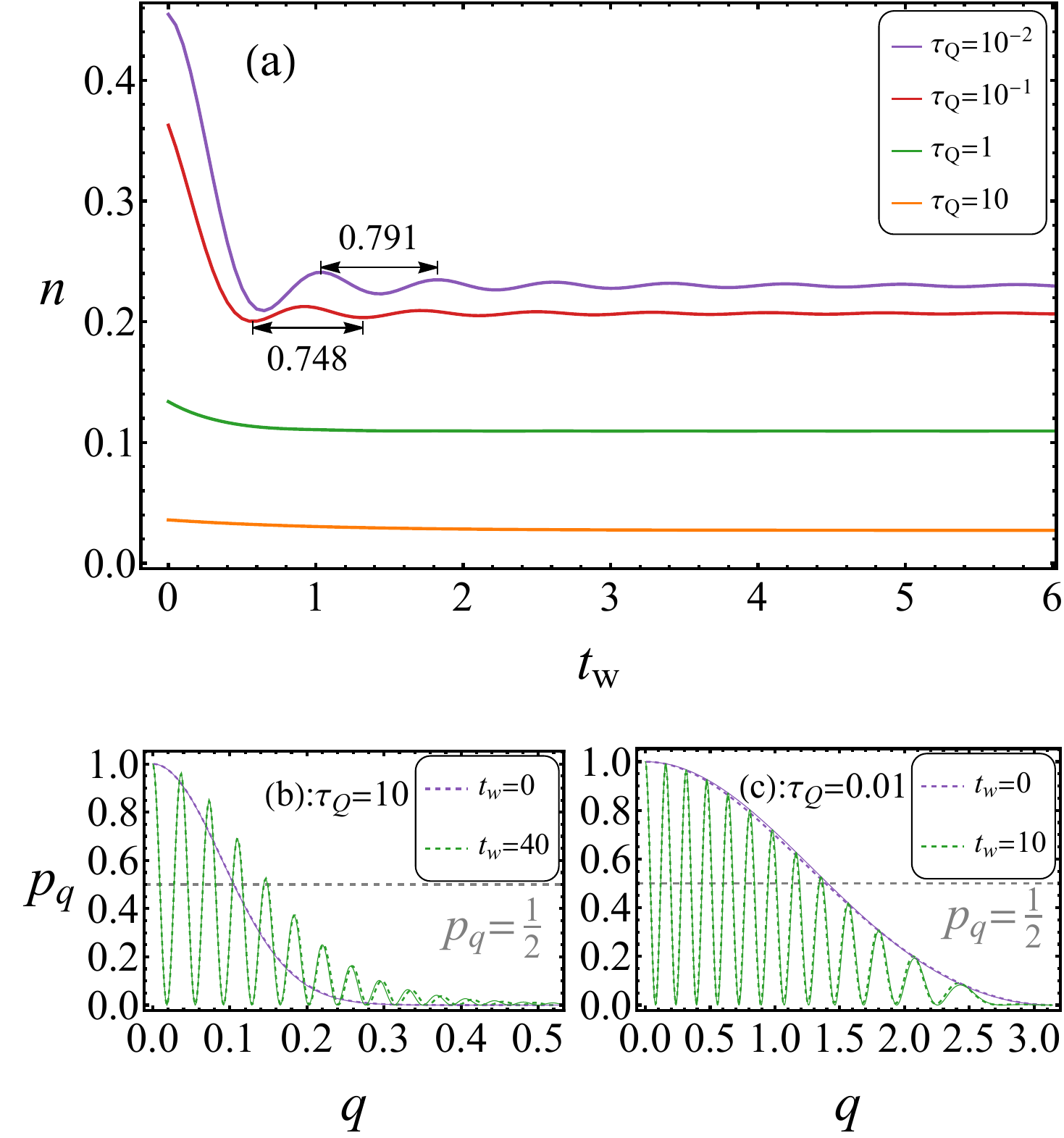}
  \end{center}
  \caption{Defect density and excitation probability modulated by the accumulated dynamical phase. (a) Defect density as a function of the waiting time for several quench times $\tau_Q$. Weakly damped oscillations persist in the fast quench, with representative periods $0.791$ for $\tau_Q=0.01$ and $0.748$ for $\tau_Q=0.1$, whereas they are strongly suppressed for slower quenches. (b) and (c) Excitation probability $p_q$. The colored dashed lines denote the numerical results. The purple and green solid lines in (b) are given by Eqs. \eqref{appeq-pqKZS} and \eqref{eq-pqKZS-waiting}, respectively. The colored solid lines in (c) are given by Eq. \eqref{eq-nsudd-waiting}. The horizontal gray dashed lines indicate $p_q=1/2$. }
  \label{plot-DD}
\end{figure}

First, the excitation probability exhibits pronounced coherent oscillations in both the slow and fast quenches. As shown in Eqs. \eqref{eq-pqKZS-waiting} and \eqref{eq-nsudd-waiting}, these oscillations originate from the phase $\phi_q^\text{w}$ accumulated during the waiting stage. Specifically, the oscillatory factor $\cos^2\phi_q^\text{w}$ indicates that the waiting stage does not merely modify the excitation amplitude, but imprints an interference pattern onto the post-quench state. Nevertheless, as shown in Fig. \ref{plot-DD}(a), the corresponding oscillations in the defect density become rapidly suppressed as the waiting time $t_w$ increases. This suppression should not be interpreted as a loss of coherence within each individual momentum. Rather, it originates from dephasing among different modes. Because different modes accumulate dynamical phases at different rates, their oscillatory contributions undergo phase averaging and destructive interference upon integration over momentum, rendering the coherent oscillations nearly invisible in the defect density. 

Compared with the slow quench, weak but discernible oscillations persist in the defect density for a fast quench. From Eq. \eqref{eq-accuDP}, the characteristic oscillation period of the defect density as a function of the waiting time $t_w$ can be estimated as $T_w\approx \pi/4$. Indeed, integrating the dynamical phase over momentum yields $\frac{1}{\pi}\int_{0}^{\pi}dq \cos^2\phi_q^{\mathrm w}=\frac{1}{2}\left[1+J_0(8t_w)\right]$, where $J_0$ denotes the Bessel function of the first kind. In the asymptotic regime $8t_w\gg1$, $J_0(8t_w)$ oscillates with an approximate period $\pi/4$ in $t_w$, thereby setting the characteristic oscillation period of the defect density.
As shown in Fig. \ref{plot-DD}(a), the numerically extracted periods are approximately $0.791$ and $0.748$ for $\tau_Q=0.01$ and $0.1$, respectively. These numerical results are close to the estimated value $\pi/4\approx 0.785$.

Second, the dynamical phase accumulated at the critical point leads to a reduction in the final defect density. Thus, relative to the $t_w=0$ protocol, the waiting stage can reduce the final density of excitations.

Third, for a slow quench, the excitation probability at $t_w=0$ takes the form $e^{-2\pi\tau_Q q^2}$, in qualitative contrast to the $t_w\rightarrow 0$ limit of Eq. \eqref{eq-pqKZS-waiting}. The origin of this difference lies in the distinct asymptotic forms of the complex parabolic cylinder functions used to evaluate $p_q$: one is appropriate in the gapped phase, whereas the other applies at the critical point. By contrast, for the fast quench, Eq. \eqref{eq-pqsudden} is naturally recovered from Eq. \eqref{eq-nsudd-waiting} by setting $t_w=0$.

Finally, for sufficiently long waiting times $t_w$ in the slow quench, Eq. \eqref{eq-pqKZS-waiting} remains quantitatively accurate only in the long-wavelength limit. This limitation arises because the phase factor $\sqrt{\pi\tau_Q}q$ is obtained within the long-wavelength approximation, and the resulting phase error becomes increasingly pronounced as $q$ increases. As shown in Fig. \ref{plot-DD}(b), the numerical results are obtained by directly solving the time-dependent Bogoliubov-de Gennes equations, whereas the purple and green curves represent the analytical predictions of Eqs. \eqref{appeq-pqKZS} and \eqref{eq-pqKZS-waiting}, respectively. For $t_w=40$, good agreement between the analytical and numerical results is found only in the vicinity of $q\rightarrow 0$. As $q$ increases, the phase discrepancy between the analytical prediction and the numerical result gradually accumulates. Nevertheless, the Gaussian envelope of Eq. \eqref{eq-pqKZS-waiting} suppresses excitations outside the long-wavelength regime, allowing the analytical expression to capture the dominant coherent oscillatory structure of the excitation probability. In contrast, for a fast quench, the analytical result agrees well with the numerical solution over almost the entire first Brillouin zone, as shown in Fig. \ref{plot-DD}(c).

\section{DQPTs with multiple critical modes}
\label{sec-dqpt}

After the two-step linear quench defined in Eq.~\eqref{eq-protocol} ($t>t_f$), the system evolves with the final Hamiltonian $H_f=H(g_f)$,
\begin{align}
  g(t)=0, ~~t>t_f.
\end{align}
The Loschmidt amplitude after the ramp factorizes as
\begin{equation}
 \mathcal{G}(\tilde{t})=\langle\psi_f|e^{-iH_f\tilde{t}}|\psi_f\rangle, ~\tilde{t}>0,
\label{eq-G}
\end{equation}
where $|\psi_f\rangle=|\psi(t_f)\rangle$ and $\tilde{t}=t-t_f$.
The Loschmidt amplitude $\mathcal{G}(\tilde{t})$ can be generalized to the complex time plane as $\mathcal{G}(z)$ (i.e., $z=\text{Re}z+i\tilde{t}$). In momentum space, it takes the form
\begin{align}\label{eq-Gzq}
  \mathcal{G}(z)=\prod_{q}\mathcal{G}_q(z)=\prod_q \left[(1-p_q)e^{\omega_q^f z}+p_q e^{-\omega_q^f z}\right],
\end{align}
where $\omega_q^f=2$. The dynamical free-energy density is defined by $f(z)=-\lim_{N\to\infty}\frac{1}{N}\ln \mathcal{G}(z)$. In the thermodynamic limit this becomes a momentum integral,
\begin{equation}
 f(z)=-\frac{1}{\pi}\int_0^\pi dq \ln\left[(1-p_q)e^{\omega_q^f z}+p_q e^{-\omega_q^f z}\right].
\label{eq-fz}
\end{equation}

\subsection{Critical Modes and Critical Times}

Nonanalyticities in $f(z)$ occur when $\mathcal{G}_q(z)=0$ (i.e., a Fisher zero).
Solving $\mathcal{G}_q(z)=0$ gives
\begin{equation}
 z_n(q)=i\frac{(2n+1)\pi}{2\omega_q^f}
 +\frac{1}{2\omega_q^f}
 \ln\frac{p_q}{1-p_q},
 \qquad n\in\mathbb{Z}.
\label{eq-fisher_zeros}
\end{equation}

The Fisher-zero line crosses the imaginary axis, corresponding to real-time evolution, only when
\begin{equation}
 p_{q_j^*}=\frac{1}{2},
\label{eq-pqhalf}
\end{equation}
where $q^*_j$ denotes the $j$-th critical mode. Here, we only discuss the range $0<q<\pi$. Each critical mode $q_j^*$ satisfying Eq.~\eqref{eq-pqhalf} corresponds to a DQPT critical time
\begin{equation}
 \tilde{t}_n^*=-i z_n(q_j^*)=\frac{(2n+1)\pi}{4}.
\label{eq-critical_times}
\end{equation}
Because $\omega_q^f=2$ is independent of momentum, all critical modes give rise to the same sequence of critical times.
Eq. \eqref{eq-pqhalf} implies that the post-quench state in the $q^*_j$ mode is an equal-amplitude superposition of the two eigenstates of $H_{q^*_j}$ (defined in Eq. \eqref{eq-hqmode}). At each critical time $t_{n}^*=t_f+\tilde{t}_{n}^*$, the evolved state becomes orthogonal to the state at $t_f$,
\begin{align}
   \langle \psi_{q^*_j}(t_f+\tilde{t}_{n}^*)|\psi_{q^*_j}(t_f)\rangle=0.
\end{align}

Consider first the case $t_w=0$, for which the protocol reduces to a one-step linear quench. The resulting DQPT is characterized by a single critical mode, located at $q^*=\sqrt{\frac{\ln 2}{2\pi\tau_Q}}$ for the slow quench and $q^*=\frac{\pi}{2}$ for the sudden quench limit with $g_i\rightarrow\infty$. As the waiting time increases, the accumulated dynamical phase $\phi_q^\text{w}$ becomes increasingly significant, leading to pronounced oscillatory structures in the excitation probability. As shown in Fig. \ref{plot-DD} (b) and (c), for $t_w\gg1$, multiple intersections of $p_q$ with the gray dashed line $p_q=1/2$ signal the emergence of an MCM regime.

We now identify the phase-interference mechanism responsible for the emergence of multiple critical modes. At the end of Sec. \ref{sec-Ising}, we compared the analytical expressions of the excitation probability with the numerical solutions of the Bogoliubov-de Gennes equations. For the slow quench, Eq. \eqref{eq-pqKZS-waiting} accurately reproduces the numerical results in the long-wavelength limit. For the fast quench, Eq. \eqref{eq-nsudd-waiting} remains quantitatively accurate over nearly the entire first Brillouin zone. 

First, for the slow quench $\tau_Q\gg1$, the coherent oscillations in Eq. \eqref{eq-pqKZS-waiting} arise from the combined phase contributions $4t_w\sin\left(\frac{q}{2}\right)$ and $\sqrt{\pi\tau_Q}q$, leading to the oscillatory factor
\begin{align}\label{eq-oscifactor}
  \cos^2\left[4t_w\sin\left(\frac{q}{2}\right)+\sqrt{\pi\tau_Q}q\right].
\end{align}
As $t_w$ increases, this phase factor oscillates rapidly as a function of momentum, resulting in a progressively shorter oscillation period in momentum space. In the long-wavelength limit, $q\rightarrow 0$, one may use $\sin(q/2)=q/2$, such that the oscillatory factor Eq. \eqref{eq-oscifactor} reduces to $\cos^2\left[(2t_w+\sqrt{\pi\tau_Q})q\right]$. The corresponding oscillation period in momentum space is therefore
\begin{align}
  T_q=\frac{\pi}{2t_w+\sqrt{\pi\tau_Q}}.
\end{align}
Meanwhile, the amplitude prefactors approach unity for $q\rightarrow 0$, $p_q^\text{half}\approx p_q^\text{su}\approx 1$ in Eqs. \eqref{eq-pqKZS-waiting} and \eqref{eq-nsudd-waiting}. The critical mode in the long-wavelength limit can thus be approximately determined from the condition Eq. \eqref{eq-pqhalf}, yielding
\begin{align}\label{eq-oscipq}
  p_{q\rightarrow0}\approx \cos^2\left[(2t_w+\sqrt{\pi\tau_Q})q\right]=\frac{1}{2}.
\end{align}
The resulting critical modes are
\begin{align} \label{slow-qj}
  q_j^*\approx\frac{(2j-1)\pi}{8t_w+4\sqrt{\pi\tau_Q}},~~j=1,2,3,\cdots,
\end{align}
where only those values of $j$ satisfying $q_j^*\rightarrow 0$ are retained. Within the long-wavelength approximation, Eq.~\eqref{slow-qj} captures the critical modes governed by the accumulated dynamical phase. 

On the other hand, for the fast quench limit $\tau_Q\rightarrow0$, the relevant dynamical phase is dominated by the waiting stage at the critical point, because of $\sqrt{\pi\tau_Q}q\rightarrow 0$ in Eq. \eqref{eq-oscifactor}. Correspondingly, the oscillation period of excitation probability and critical modes of DQPT are $T_q=\frac{\pi}{2t_w}$ and $q_j^*\approx\frac{(2j-1)\pi}{8t_w}$, respectively. For both the slow and fast quenches, the spacing between adjacent critical modes approximately scales as $1/t_w$. Consequently, increasing the waiting time enhances the phase accumulation and thereby yields an increasing number of critical modes. Thus, the waiting time directly controls the number of critical modes and can drive the DQPT from the SCM regime to the MCM regime.

\begin{figure}[t]
  \begin{center}
		\includegraphics[width=3.2 in,angle=0]{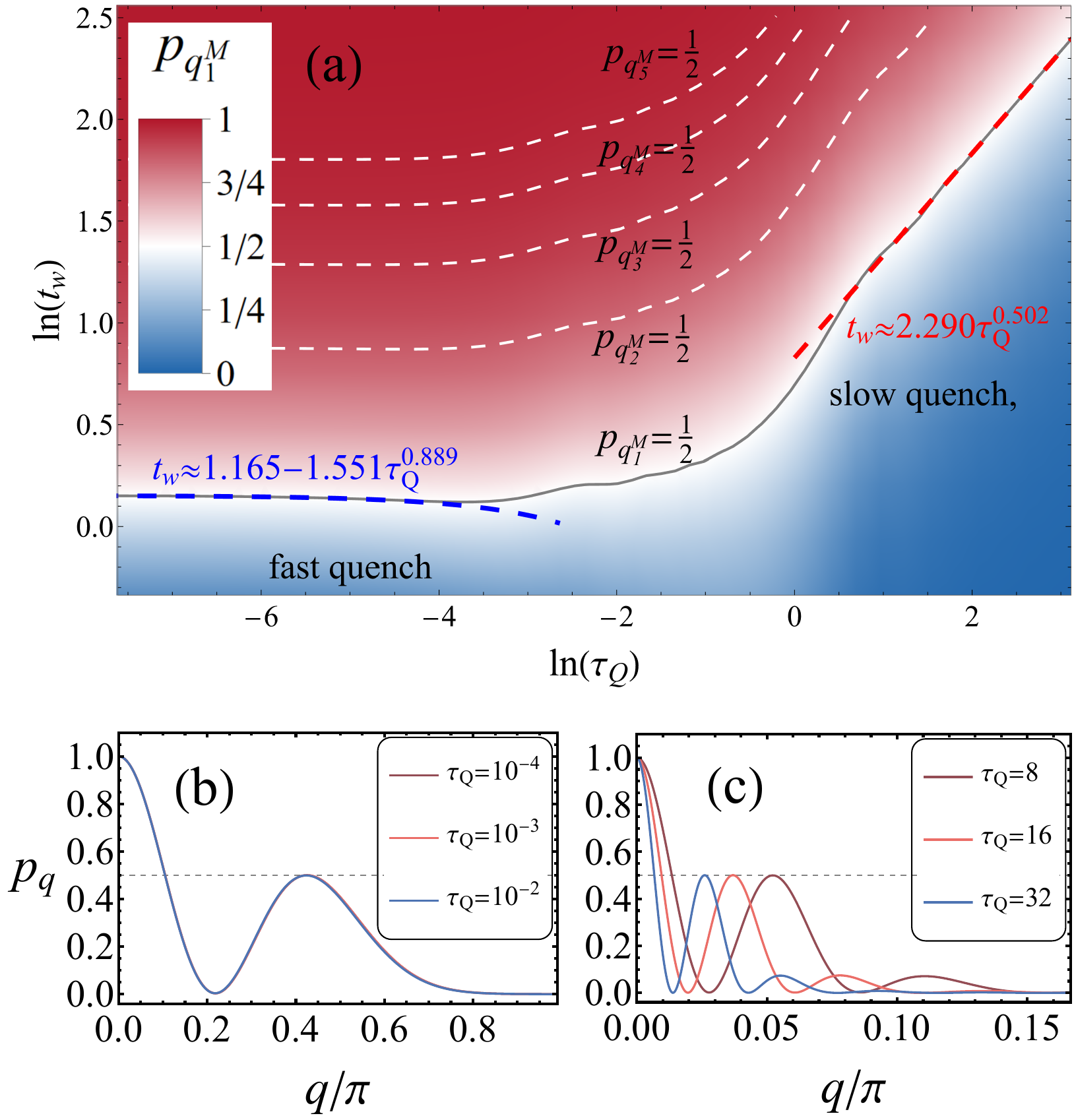}
  \end{center}
  \caption{ (a) Density plot of $p_{q_1^\text{M}}$ as a function of $\tau_Q$ and $t_w$ for $g_i=6$, where $q_1^\text{M}$ denotes the momentum corresponding to the first local maximum of $p_q$ within $0<q<\pi$. The blue and red regions correspond to the SCM and MCM regimes, respectively. The gray contour, defined by $p_{q_1^\text{M}}=1/2$, separates the SCM and MCM regimes. The blue and red dashed lines show the asymptotic relations between $t_w$ and $\tau_Q$ along this boundary for the fast and slow quenches, respectively. Within the MCM regime, the white dashed lines denote $p_{q_j^\text{M}}=1/2$ with $j=2,3,4,5$. (b) and (c) Excitation probability $p_q$ at the SCM--MCM boundary for representative values of $\tau_Q$ and $t_w$. The horizontal dashed line indicates $p_q=1/2$.}
  \label{plot-pqM1}
\end{figure}

\subsection{Boundary between SCM and MCM regimes}
\label{subsec-boundary}

We now extend the analysis from the long-wavelength limit to the positive half of the first Brillouin zone, $0<q<\pi$. Except for the critical mode $q_1^*$, the remaining critical modes are generally created or annihilated in pairs. As shown in Figs. \ref{plot-DD} (b) and (c), the coherent oscillations induced by the dynamical phase generate a sequence of local maxima in the excitation probability $p_q$. We denote the momentum of the $m$-th local maximum by $q_m^\text{M}$. Within the positive half of the first Brillouin zone, their positions are ordered as
\begin{align}
  0<q_1^\text{M}<q_2^\text{M}<q_3^\text{M}<\cdots<\pi,
\end{align}
For the parameter regime considered here, the heights of these successive maxima decrease monotonically,
\begin{align}
  1>p_{q_1^\text{M}}>p_{q_2^\text{M}}>p_{q_3^\text{M}}>\cdots>0,
\end{align}
When a given local maximum satisfies $p_{q_m^\text{M}}>1/2$, the excitation probability curve crosses the line $p_q=1/2$ on both sides of the maximum, giving rise to a pair of critical modes, denoted by $q_{2m}^*$ and $q_{2m+1}^*$. By contrast, when $p_{q_m^\text{M}}<1/2$, no critical mode exists in the vicinity of that maximum. 
Accordingly, $p_{q_m^\text{M}}=1/2$ marks the creation or annihilation threshold of a pair of critical modes. At this condition, the two critical modes merge at $q_m^{\mathrm M}$, forming a tangential critical mode at which the excitation probability curve is tangent to the line $p_q=1/2$:
\begin{align}\label{eq-pqmMmcm}
  p_{q_m^{\mathrm M}}=\frac{1}{2}~~\text{and}~~\left.\frac{dp_q}{dq}\right|_{q\rightarrow q_m^{\mathrm M}}=0.
\end{align} 
As shown below, the coalescence converts a linear crossing of $p_q=1/2$ into a quadratic tangency. This change in the local momentum dependence preserves the nonanalyticity of the rate function but changes its critical exponent.

For sufficiently short waiting times $t_w$, we have $p_{q_1^\text{M}}<1/2$. In this case, only the critical mode $q_1^*$ is present, and the DQPT is characterized by the SCM regime. As $t_w$ increases, the coherent oscillations induced by the dynamical phase become increasingly pronounced. Once the largest local maximum of excitation probability $p_{q_1^\text{M}}$ exceeds 1/2, an additional pair of critical modes emerges, and the system enters the MCM regime. Consequently, the SCM--MCM boundary is determined by
\begin{align} \label{eq-pq1Mhalf}
  p_{q_1^\text{M}}=1/2.
\end{align}
Fig. \ref{plot-pqM1}(a) shows the density plot of $p_{q_1^\text{M}}$ as a function of the quench time $\tau_Q$ and the waiting time $t_w$ for $g_i=6$. The blue region corresponds to the SCM regime, whereas the red region represents the MCM regime. The two regimes are separated by the gray contour defined by Eq. \eqref{eq-pq1Mhalf}.

For the fast quench, the analytical expression for the excitation probability, Eq. \eqref{eq-nsudd-waiting}, indicates that the waiting time approaches a constant
\begin{align}
  t_w \rightarrow C,~~\tau_Q\rightarrow0, \label{eq-crttw-fast}
\end{align}
when $p_{q_1^\text{M}}=1/2$. A numerical fit to the boundary, as indicated by the blue dashed line in Fig. \ref{plot-pqM1}(a), yields $t_w\approx 1.165-1.551\tau_Q^{0.889}$. In the limit $\tau_Q\rightarrow 0$, the waiting time is a finite value $t_w\approx1.165$. For the slow quench, $p_q$ asymptotically exhibits a Gaussian envelope in the long-wavelength limit, $p_q^\text{half}\approx e^{-a \tau_Q q^2}$. As a result, when $p_{q_1^\text{M}}=1/2$, we have $p_{q_1^\text{M}}^\text{half}\approx 1/2$ and $\cos^2(\phi^\text{w}_{q_1^\text{M}}+\sqrt{\pi\tau_Q}q_1^\text{M})\approx1$. The former gives the relation between $q_1^\text{M}$ and $\tau_Q$, $q_1^\text{M}\propto 1/\sqrt{\tau_Q}$. The latter gives $(2t_w+\sqrt{\pi\tau_Q})q_1^\text{M}\approx \pi$. Combining these two relations yields the square-root scaling
\begin{align}
  t_w\propto\sqrt{\tau_Q},~~\tau_Q\gg1.  \label{eq-crttw-slow}
\end{align}
This square-root law follows from the KZ scaling encoded in the asymptotic Gaussian envelope $p_q^\text{half}$. As shown by the red dashed line in Fig. \ref{plot-pqM1}(a), a numerical fit to the boundary gives $t_w\approx2.290~\tau_Q^{0.502}$ providing strong numerical support for the square-root scaling.

Figs. \ref{plot-pqM1}(b) and (c) depict the excitation probabilities at the SCM–MCM boundary in the fast and slow quenches, respectively. In both cases, Eq. \eqref{eq-pq1Mhalf} holds. Finally, since the crossover from the fast quench to the slow quench involves an intermediate region \cite{LiP_2023}, neither asymptotic fitting form provides an accurate description throughout the crossover. Consequently, no single simple functional relation between $t_w$ and $\tau_Q$ can be extracted in the vicinity of $\ln\tau_Q\approx 0$.

\section{Rate function and critical behavior}
\label{sec-ratefunc}

\subsection{Rate Function and Dynamical Winding Number}
In the thermodynamic limit, the real-time Loschmidt rate function is defined as
\begin{equation}
  r(\tilde{t})=-\lim_{N\rightarrow\infty}\frac{1}{N}\ln |\mathcal{G}(\tilde{t})|^2 .
\label{eq-rate-def}
\end{equation}
Using Eq. \eqref{eq-Gzq}, this rate function becomes
\begin{equation}
  r(\tilde{t})=-\frac{1}{\pi}\int_0^\pi d q\,
     \ln\left[1-4p_q(1-p_q)\sin^2(2 \tilde{t})\right].
\label{eq-rate}
\end{equation}
Eq.~\eqref{eq-rate} shows directly why the condition $p_q=1/2$ is special \cite{Polkovnikov_2016}. When a critical mode $q_j^*$ satisfies $p_{q_j^*}=1/2$, the rate function can become singular at the critical time $\tilde{t}_{n}^*=(2n+1)\pi/4$. Because $\omega_q^f=2$, all critical modes generate singularities in $r(\tilde{t})$ at the same sequence of critical times.

We next examine the singular behavior of the rate function near the critical times. At a critical time $\tilde{t}_{n}^*$, the rate function reaches a finite but nonanalytic value $r(\tilde{t}_{n}^*)$. Its local scaling behavior in the vicinity of the critical time is characterized by the critical exponent $\alpha$ \cite{Hely_2015PRL, WuYT_2020},
\begin{align}
  r(\tilde{t})-r(\tilde{t}_{n}^*)\propto |\tilde{t}-\tilde{t}_{n}^*|^\alpha.
  \label{eq-rtexpand}
\end{align}
The critical exponent $\alpha$ can be inferred from the local behavior of the excitation probability near the corresponding critical mode.

We first consider $q_1^*$ and the other ordinary critical modes that occur in pairs. In this case, $p_q$ varies monotonically in the vicinity of $q_j^*$ and crosses $p_{q_j^*}=1/2$ with a nonzero slope. It can therefore be expanded as $p_q=1/2-a_j(q-q_j^*)$, where $a_j$ is a nonzero coefficient. By substituting this expansion into the rate function and evaluating the leading contribution around $q_j^*$, we find the critical exponent $\alpha=1$, with the details given in App. \ref{app-criticalE}. From Eq. \eqref{eq-rtexpand}, the first derivative of the rate function remains finite on both sides of the critical time but approaches different left and right limits,
\begin{align}
  \frac{dr(\tilde{t})}{d\tilde{t}}\propto \text{sgn}(\tilde{t}-\tilde{t}_{n}^*)|\tilde{t}-\tilde{t}_{n}^*|^0.
\end{align}
Thus, an ordinary critical mode produces a linear cusp in the rate function, accompanied by a finite discontinuity in its first derivative.

At the SCM--MCM boundary, two critical modes coalesce at the first local maximum of the excitation probability and form a tangential critical mode, where $q_1^\text{M}=q_2^*=q_3^*$ and $p_{q_1^\text{M}}=1/2$. Near $q_1^\text{M}$, $p_q$ can be expanded as $p_q=1/2-b_1(q-q_1^\text{M})^2$, where $b_1$ is a nonzero coefficient. Because the linear crossing is replaced by this quadratic tangency, substitution into the rate function and integration around $q_1^\text{M}$ yield the distinct exponent $\alpha=1/2$. The rate function therefore exhibits a square-root singularity near the critical time, and its first derivative diverges as
\begin{align}
  \frac{dr(\tilde{t})}{d\tilde{t}}\propto \text{sgn}(\tilde{t}-\tilde{t}_{n}^*)|\tilde{t}-\tilde{t}_{n}^*|^{-\frac{1}{2}}.
\end{align}

The same $\alpha=1/2$ singularity is not restricted to the SCM--MCM boundary. Within the MCM regime, the quench time $\tau_Q$ and waiting time $t_w$ can be tuned such that other local maxima satisfy Eq. \eqref{eq-pqmMmcm}, as shown in the white dashed lines in Fig. \ref{plot-pqM1}. Under this condition, a pair of critical modes coalesces at the mode $q_m^\text{M}$, where $p_q$ can be expanded as $p_q-1/2\propto (q-q_m^\text{M})^2$ and the critical structure of the rate function is likewise characterized by $\alpha=1/2$. Therefore, under the quench protocol defined in Eq. \eqref{eq-protocol}, DQPT critical behavior with $\alpha=1/2$ can be realized by tuning the quench time and waiting time.

The geometric phase provides a complementary characterization of these DQPTs.
For each momentum sector, the Loschmidt amplitude after the ramp is
\begin{align}\label{eq-losAmp-q}
  \mathcal{G}_q(\tilde{t})=\cos(2 \tilde{t})+i(1-2p_q)\sin(2 \tilde{t}).
\end{align}
The Pancharatnam phase $\phi_q(\tilde{t})=\arg \mathcal{G}_q(\tilde{t})$ reads
\begin{equation}
    \phi_q(\tilde{t})
    =\arg\left[\cos(2 \tilde{t})+i(1-2p_q)\sin(2 \tilde{t})\right].
\label{eq-pancharatnam-phase}
\end{equation}
The accumulated dynamical phase under the final Hamiltonian is
\begin{equation}
   \phi_q^{\rm dyn}(\tilde{t})=(1-2p_q)2 \tilde{t}.
\label{eq-dyn-phase}
\end{equation}
The Pancharatnam geometric phase is
\begin{equation}
     \phi_q^{\rm G}(\tilde{t})=\phi_q(\tilde{t})-\phi_q^{\rm dyn}(\tilde{t}).
\label{eq-geometric-phase}
\end{equation}
At a critical mode $q_j^*$ with $p_{q_j^*}=1/2$, Eq.~\eqref{eq-losAmp-q} reduces to $\mathcal{G}_{q_j^*}(\tilde{t})=\cos(2 \tilde{t})$. 

\begin{figure}[t]
  \begin{center}
        \includegraphics[width=3.0 in,angle=0]{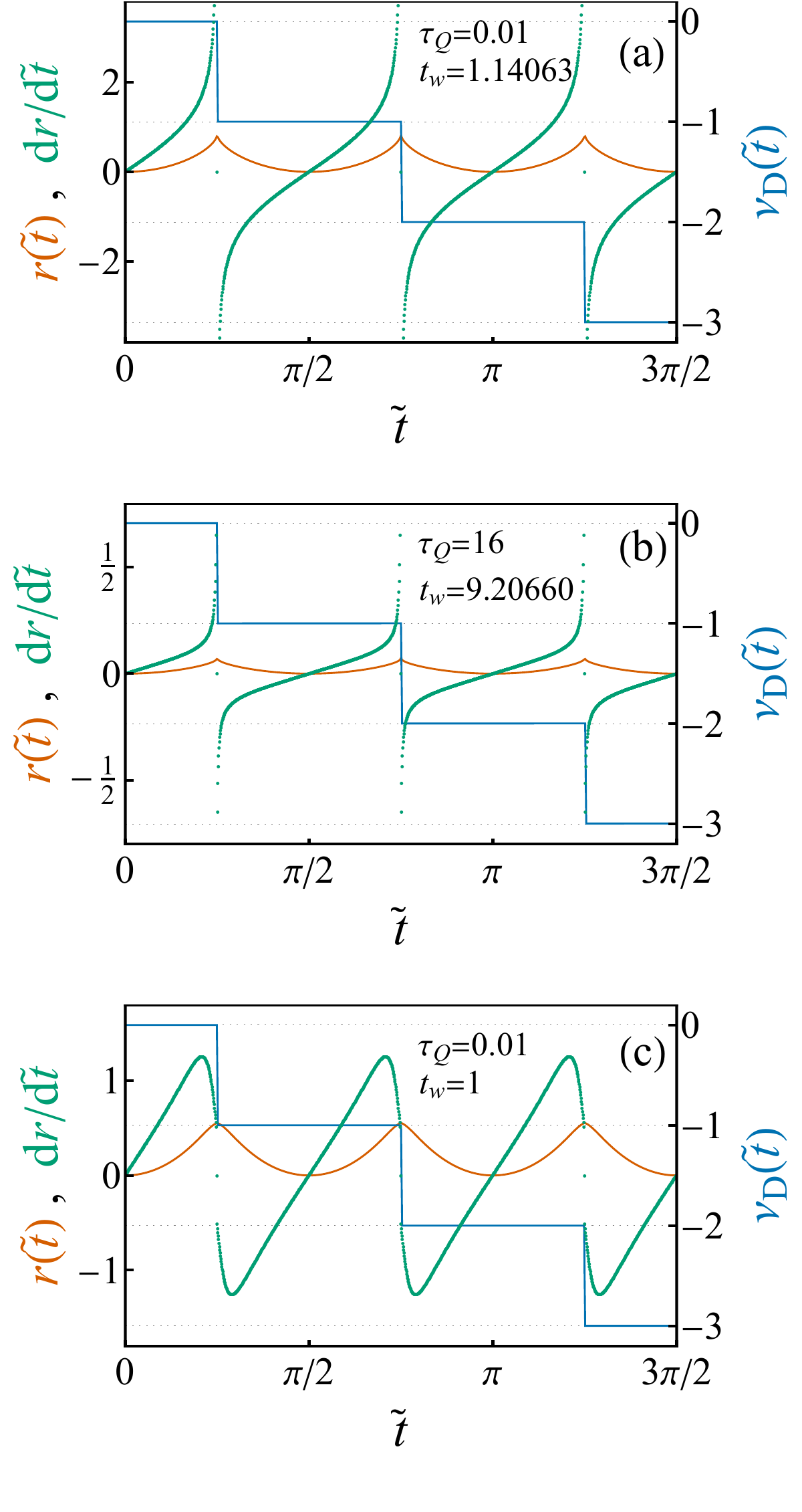}
  \end{center}
  \caption{ Rate function $r(\tilde{t})$, its derivative $\dot{r}(\tilde{t})$, and dynamical winding number $\nu_D(\tilde{t})$ of the post-quench state for $g_i=6$. In (a) and (b), the quench parameters $(\tau_Q,t_w)$ are chosen on the SCM–MCM boundary, whereas in (c) they are chosen in the SCM regime. At each critical time $\tilde{t}_n^{*}$, $r(\tilde{t})$ is nonanalytic and $\nu_D(\tilde{t})$ exhibits a jump. On the SCM–MCM boundary, as shown in (a) and (b), $\dot{r}(\tilde{t})$ diverges as $|\tilde{t}-\tilde{t}_n^{*}|^{-1/2}$, whereas in the SCM regime, as shown in (c), $\dot{r}(\tilde{t})$ exhibits a finite discontinuity.
  }
  \label{plot-rate}
\end{figure}

We quantify this geometric phase change by the dynamical winding number
\begin{equation}
 \nu_D(\tilde{t})=\frac{1}{2\pi}\int_0^\pi \partial_q \phi_q^{\rm G}(\tilde{t})\,d q.
\label{eq-nuD}
\end{equation}
At a DQPT critical time, a conventional critical mode $q^*_j$ renders the geometric phase undefined and induces a quantized jump in $\nu_D(\tilde{t})$. By contrast, a tangential critical mode $q_j^\text{M}$ leaves $\nu_D(\tilde{t})$ unchanged.

\subsection{Numerical Results}

In Sec. \ref{subsec-boundary}, we analytically and numerically determined the SCM--MCM boundary in the $\tau_Q-t_w$ parameter plane. Here, we present the critical structure of DQPTs on the boundary and compare it with that in the SCM regime.

Fig. \ref{plot-rate} shows the rate function $r(\tilde{t})$ (orange curves), its first derivative $dr/d\tilde{t}$ (green dotted curves), and the dynamical winding number $\nu_\text{D}(\tilde{t})$ (blue solid lines). The numerical values of $r(\tilde{t})$ and $\nu_\text{D}(\tilde{t})$ are obtained by discretizing the momentum integrals in Eqs. \eqref{eq-rate} and \eqref{eq-nuD} into finite sums over the momenta in the first Brillouin zone. The first derivative of the rate function can be expressed as
\begin{equation}
  \frac{dr(\tilde{t})}{d\tilde{t}}=\frac{1}{\pi}\int_0^\pi d q\,
     \frac{4p_q(1-p_q)\omega_q^f\sin(2\omega_q^f \tilde{t})}{1-4p_q(1-p_q)\sin^2(\omega_q^f \tilde{t})}.
\label{eq-drate}
\end{equation}
It is evaluated numerically using the same discrete momentum summation. Figs. \ref{plot-rate}(a) and (b) correspond to two representative points on the SCM--MCM boundary, with $(\tau_Q,t_w)=(0.01,1.14063)$ and $(16,9.20660)$, respectively. Fig. \ref{plot-rate}(c) shows a representative point in the SCM regime, with $(\tau_Q,t_w)=(0.01,1)$.

As shown in Fig. \ref{plot-rate}, the dynamical winding number exhibits quantized jumps. These jumps occur at the critical times given by Eq. \eqref{eq-critical_times}. On the SCM--MCM boundary, the rate function is nonanalytic at the critical time and approaches the nonanalytic point with a nonlinear dependence on $|\tilde{t}-\tilde{t}_n^*|$, while its first derivative $dr/d\tilde{t}$ diverges. In contrast, in the SCM regime, the rate function approaches the nonanalytic point linearly, and $dr/d\tilde{t}$ remains finite but discontinuous. More generally, within the MCM regime, the divergence of $dr/d\tilde{t}$ occurs whenever $p_{q_j^M}=1/2$, corresponding to the white dashed lines shown in Fig. \ref{plot-pqM1}(a). Away from these curves, $dr/d\tilde{t}$ remains finite but discontinuous at the critical times.

\begin{figure}[t]
  \begin{center}
		\includegraphics[width=3.3 in,angle=0]{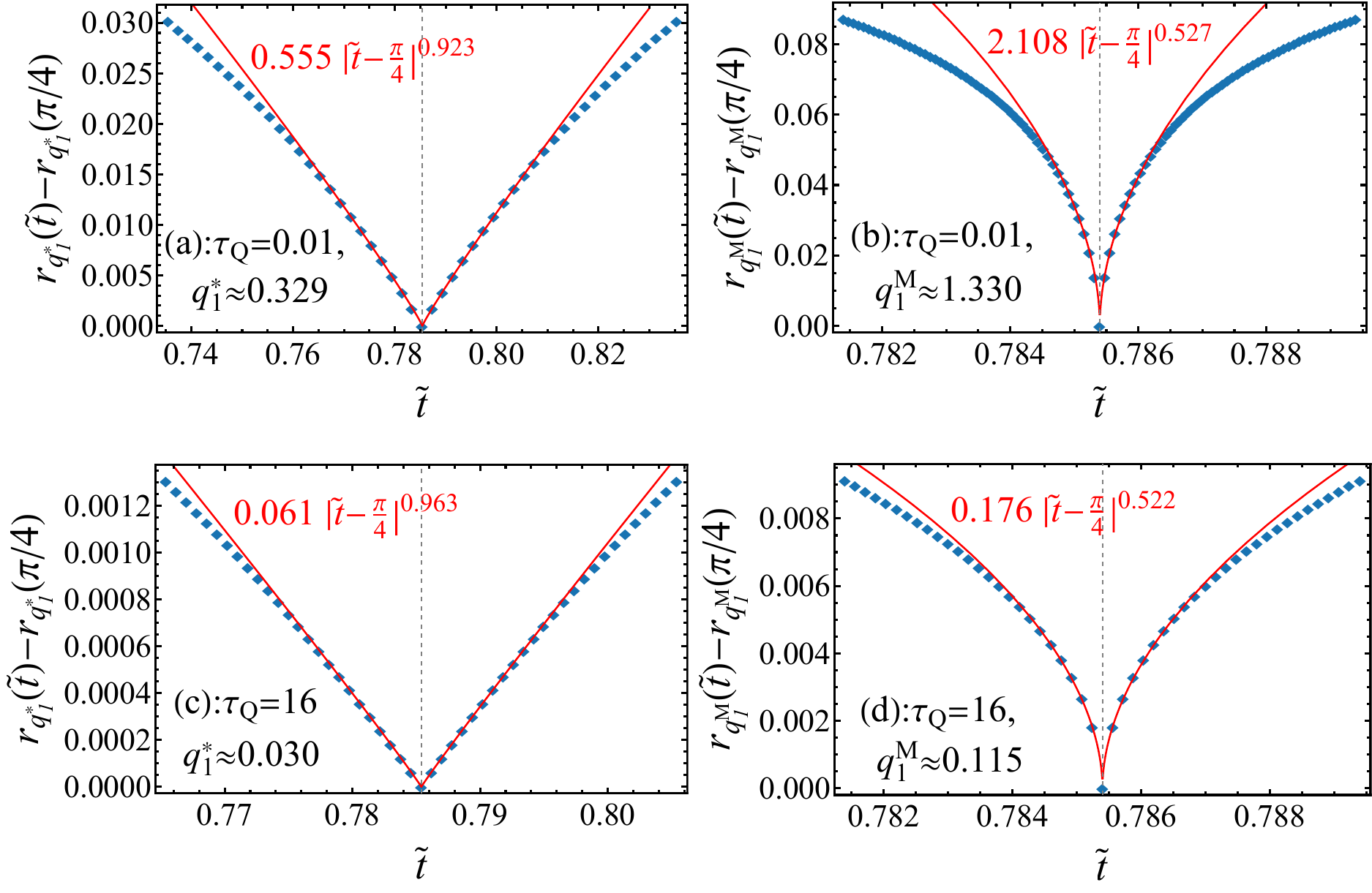}
  \end{center}
  \caption{ Scaling behavior of the critical-mode rate function near the critical time $\tilde{t}^*_0=\pi/4$. The blue symbols represent the numerical results, while the red solid lines denote the fitting result of $r_{\bar{q}}(\tilde{t})-r_{\bar{q}}(\pi/4)$. The vertical gray dashed lines mark a critical time $\tilde{t}^*_0=\pi/4$. Panels (a) and (b) correspond to $\bar{q}=q_1^*\approx 0.329$ and $\bar{q}=q_1^{\mathrm M}\approx 1.330$, respectively, for the fast quench with $\tau_Q=0.01$ and $t_w=1.14063$. The fitted exponents are $0.923$ and $0.527$. Panels (c) and (d) show the corresponding results for the slow quench with $\tau_Q=16$ and a waiting time $t_w=9.20660$, with $q_1^*\approx 0.030$ and $q_1^{\mathrm M}\approx 0.115$, yielding the fitted exponents $0.963$ and $0.522$, respectively. Thus, the contributions from $q_1^*$ and $q_1^{\mathrm M}$ exhibit critical exponents $\alpha\approx 1$ and $\alpha \approx 1/2$, respectively. }
  \label{plot-criticalE}
\end{figure}

On the SCM--MCM boundary, the post-quench system possesses both an ordinary critical mode $q_1^*$ and a tangential critical mode $q_1^\text{M}$, with the latter arising from the coalescence of a pair of critical modes. Although the two critical modes generate the same sequence of critical times, the associated nonanalyticities of the rate function are characterized by distinct critical exponents $\alpha$. To disentangle the contributions associated with different critical modes, we introduce a critical-mode rate function $r_{\bar{q}}(\tilde{t})=-(1/\pi)\int_{\bar{q}-\delta}^{\bar{q}+\delta}dq \ln|\mathcal{G}_q(\tilde{t})|^2$, where $\bar{q}$ is chosen as either of the two critical modes, $q_1^*$ or $q_1^\text{M}$, with $0<\delta\ll1$ chosen to define a narrow momentum window around $\bar{q}$. Accordingly, $r_{\bar{q}}(\tilde{t})$ isolates the local contribution of the critical mode $\bar{q}$ to the full rate function. After subtracting its value at the critical time $\tilde{t}^*_0=\pi/4$, the leading nonanalytic behavior is characterized by $\Delta r_{\bar{q}}(\tilde{t})\equiv r_{\bar{q}}(\tilde{t})-r_{\bar{q}}(\pi/4)\propto |\tilde{t}-\pi/4|^{\alpha_{\bar{q}}}$. As shown in Fig. \ref{plot-criticalE}, panels (a) and (b) display the contributions associated with $q_1^*$ and $q_1^\text{M}$ respectively, for the fast quench with $\tau_Q=0.01$. The fitted exponents are $0.923$ and $0.527$, corresponding to $\alpha_{q_1^*}\approx 1$ and $\alpha_{q_1^\text{M}}\approx 1/2$, respectively. Panels (c) and (d) show the corresponding results for the slow quench with $\tau_Q=16$, yielding the fitted exponents $0.963$ and $0.522$. These results consistently demonstrate that the ordinary critical mode $q_1^*$ gives rise to an approximately linear cusp, whereas the tangential critical mode $q_1^\text{M}$, located at a local maximum of the excitation probability, produces a square-root nonanalyticity.

The rate function near the critical time contains two power-law contributions, proportional to $|\tilde{t}-\pi/4|$ and $|\tilde{t}-\pi/4|^{1/2}$, respectively. The square-root contribution dominates as $\tilde{t}\rightarrow \pi/4$. Consequently, the first derivative of the rate function diverges at the critical time. By contrast, the quantized jump in the dynamical winding number is generated by the ordinary critical mode $q_1^*$. Although the tangential critical mode $q_1^\text{M}$ modifies the local singular structure of the rate function, it does not produce an additional jump in the dynamical winding number.

Notably, in Figs. \ref{plot-rate} and \ref{plot-criticalE}, the quoted values of $t_w$ are specified to five decimal places so that the tangency condition $p_q=1/2$ at $q_1^\text{M}$ is satisfied to the numerical precision used in the scaling analysis.

\section{Extended SSH model} \label{sec-ESSH}

\begin{figure}[t]
  \begin{center}
		\includegraphics[width=3.0 in,angle=0]{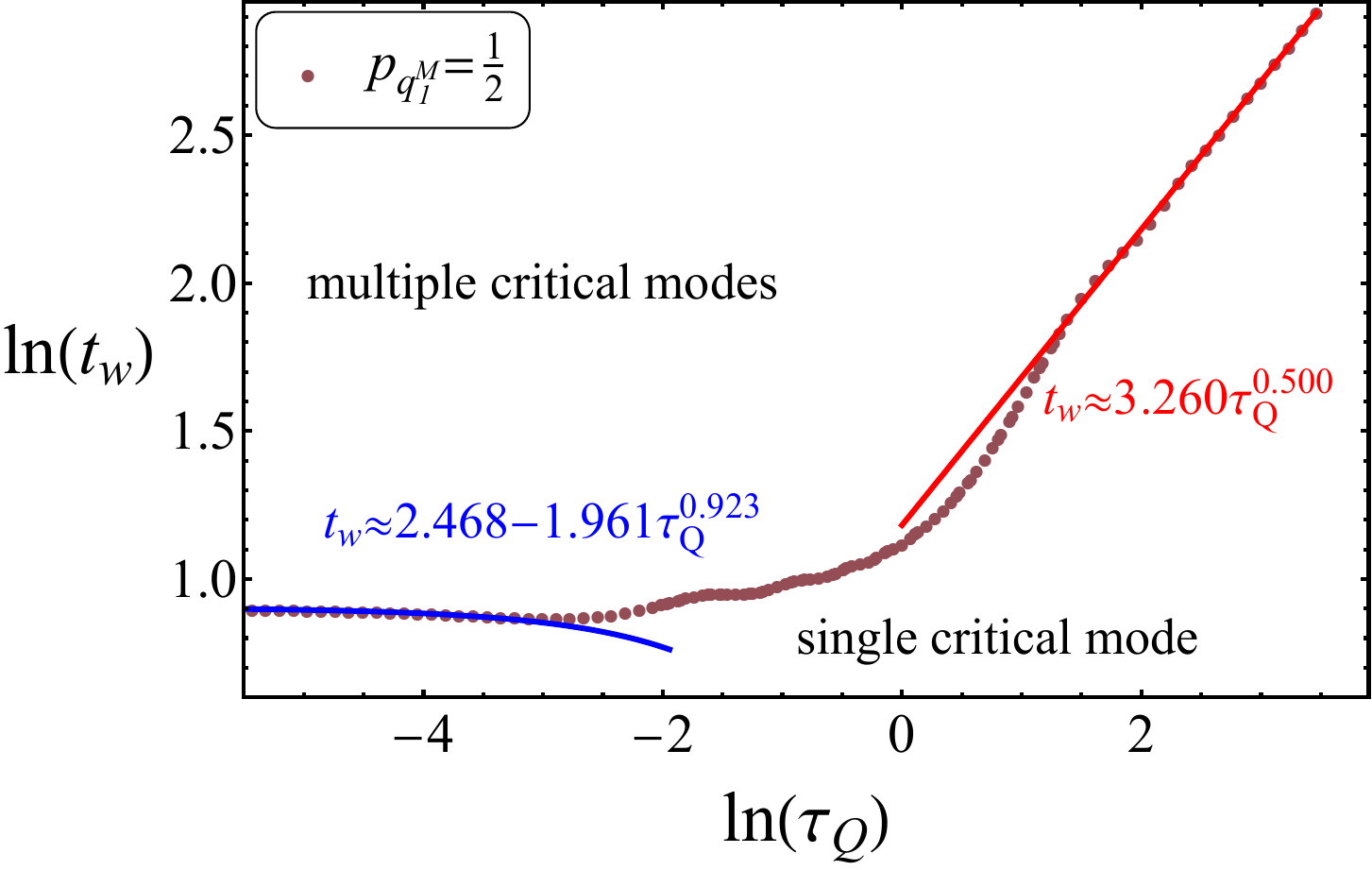}
  \end{center}
  \caption{Boundary between the SCM and MCM regimes in the extended SSH model for $J_i=6$. The blue and red lines show the asymptotic relations between $t_w$ and $\tau_Q$ along this boundary in the fast and slow quenches, respectively.}
  \label{plot-SSHpqm1}
\end{figure}

To demonstrate that our findings are not limited to the transverse Ising chain, we next consider an extended SSH model exhibiting a topological quantum phase transition,
\begin{eqnarray}
  &H=\sum_n\left(J_1 a_n^\dagger b_n+J_2 a_{n+1}^\dagger b_n +J_3  a_{n+2}^\dagger b_n+\text{H}.\text{c}.\right),~~~
\end{eqnarray}
where $J_1$, $J_2$ and $J_3$ describe intracell, nearest-neighbor intercell, and next-nearest-neighbor intercell hopping amplitudes. By Fourier transformation, the SSH model can be decomposed into a series of two-level systems $H_q=d_x^q\sigma^x+d_y^q\sigma^y$ where $\sigma^s$ ($s=x, y ,z$) are the Pauli matrices, and
\begin{align}
  d_q^x&=J_1+J_2\cos q+J_3\cos 2q,\\
  d_q^y&=J_2\sin q+J_3\sin 2q.
\end{align}
The spectrum is $\omega_\pm=\pm[(d_q^x)^2+(d_q^y)^2]^{1/2}$. We set $J_1=1$ as the unit of energy and $J_2=J_3=-\frac{1}{2}J$. Following the protocol introduced in Eq. \eqref{eq-protocol}, we apply to the present model a two-step linear quench with a waiting stage at the critical point $J_c=1$,
\begin{align} \label{eq-SSHprotocol}
 J(t)=\left\{
 \begin{array}{cc}
 J_i-\frac{t}{\tau_Q},
 & 0\leq t<t_1,\\
 J_c,
 & t_1 \leq t\leq t_2,\\
 J_i-\frac{t-t_w}{\tau_Q},
 &  t_2 \leq t < t_f,
 \end{array}\right.
\end{align}
where $t_1=(J_i-J_c)\tau_Q$, $t_2=t_1+t_w$, and $t_f= J_i\tau_Q+t_w$. At $t_f$, the intercell hopping amplitudes vanish, $J(t_f)=0$.

By introducing a transformation matrix
\begin{align}
  S=\frac{1}{\sqrt{2}}\left(
  \begin{array}{cc}
    e^{-i3q/2}&-e^{-i3q/2} \\
    1 & 1 
  \end{array}
  \right),
\end{align}
the time-dependent parameter $J(t)$ appears in the diagonal part of the Hamiltonian. The transformed Hamiltonian reads
\begin{align}
  H'_q&=S^\dagger H_qS\nonumber\\
  &=\left[\cos\frac{3q}{2}-J(t)\cos\frac{q}{2}\right]\sigma^z-\sin\frac{3q}{2}\sigma^y.
\end{align}

\begin{figure}[t]
  \begin{center}
        \includegraphics[width=3.45 in,angle=0]{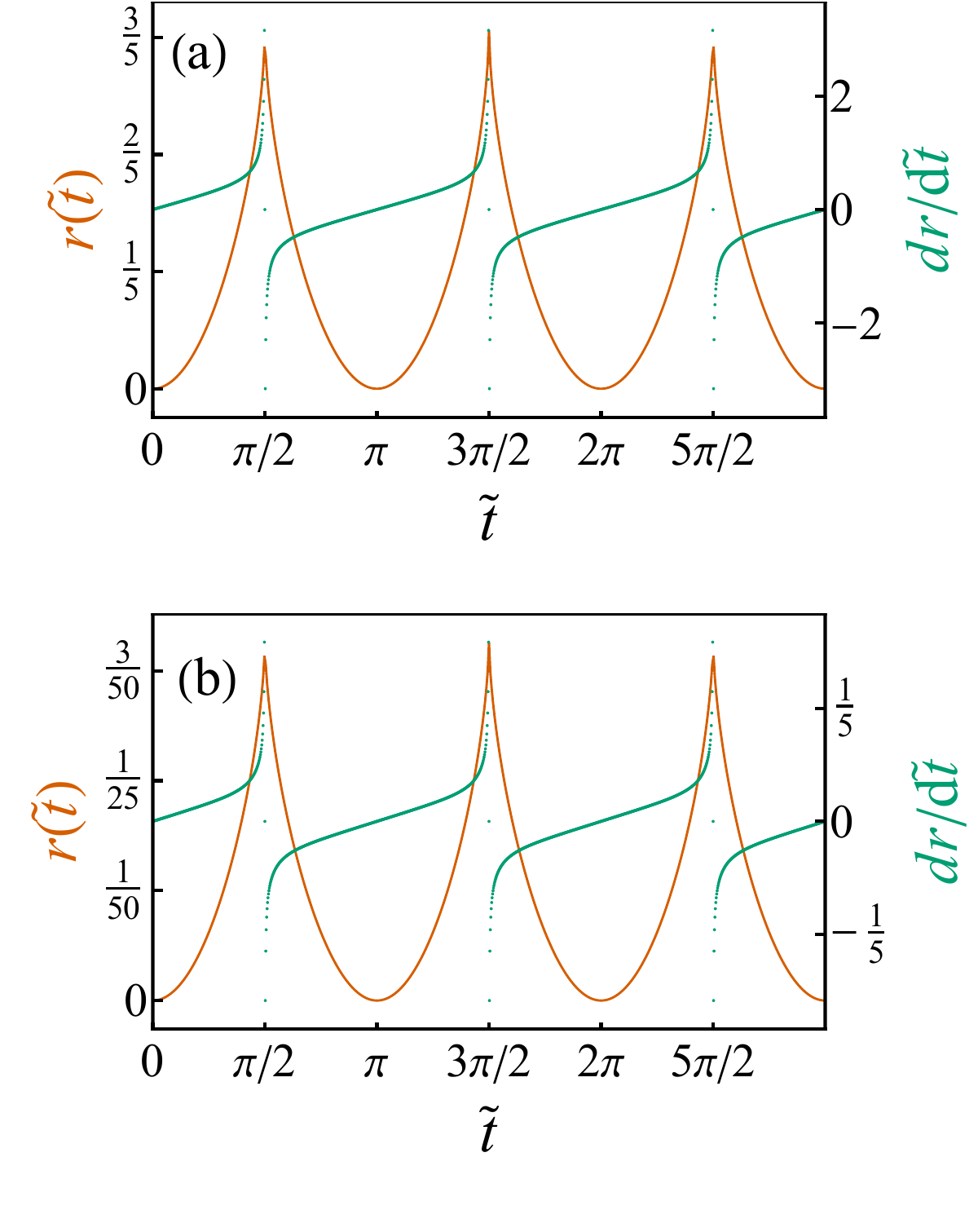}
  \end{center}
  \caption{$r(\tilde{t})$ and $\dot{r}(\tilde{t})$ after the ramp with (a) $\tau_Q=0.01$ and $t_w=2.44057$ and (b)  $\tau_Q=16$ and $t_w=13.04719$. The parameter is fixed at $J_i=6$. At the critical time $\tilde{t}^*_{n}$, $r(\tilde{t})$ is nonanalytic, and $\dot{r}(\tilde{t})$ diverges as $|\tilde{t}-\tilde{t}^*_{n}|^{-1/2}$.}
  \label{plot-SSHrateF}
\end{figure}

When the quench protocol is completed, the rate function of the post-quench state shows nonanalytic behavior at each critical time $t_{n}^*=t_f+(2n+1)\pi/2$ or $\tilde{t}_{n}^*=t_{n}^*-t_f=(2n+1)\pi/2$ (with $n=0,1,2,3,\cdots$). The rate function reads
\begin{equation}
  r(\tilde{t})=-\frac{1}{\pi}\int_0^\pi d q\,
     \ln\left[1-4p_q(1-p_q)\sin^2\tilde{t}~\right],
\end{equation}
where the excitation probability $p_q$ exhibits coherent oscillations induced by the accumulated dynamical phase at the critical point.

As in the Ising chain, the oscillations induced by the dynamical phase generate additional critical modes. Fig. \ref{plot-SSHpqm1} shows the corresponding SCM--MCM boundary, on which we have $p_{q_1^\text{M}}=1/2$. In the fast quench limit, the waiting time at this boundary approaches a constant, consistent with the result for the Ising model in Eq. \eqref{eq-crttw-fast}. By contrast, in the slow quench case, the dependence of the waiting time on $\tau_Q$ follows the KZ scaling given in Eq. \eqref{eq-crttw-slow}. The numerical fits shown by the red and blue lines in Fig. \ref{plot-SSHpqm1} provide strong support for these asymptotic predictions. 

Finally, in Fig. \ref{plot-SSHrateF}, we exhibit the rate function and its first derivative for a fast quench with $\tau_Q=0.01$ and a slow quench with $\tau_Q=16$. The corresponding values of $t_w$ are $2.44057$ and $13.04719$, respectively. Near each critical time, the rate function approaches the nonanalytic point with a square-root form, and its first derivative diverges. This singular behavior indicates that the critical exponent of the rate function is no longer the conventional value $\alpha=1$, but instead becomes $\alpha=1/2$.

\section{Conclusion}
\label{sec-conclusion}

In summary, we have studied DQPTs in the transverse Ising chain and the extended SSH model under a two-step linear quench containing a finite waiting stage at an equilibrium critical point. Our results reveal that the dynamical phase accumulated at the critical point plays a key role in controlling the critical structure of DQPTs. The accumulated phase produces coherent oscillations in the excitation probability, allowing the system to evolve from the SCM regime to the MCM regime as the waiting time is increased. We determined this boundary in the fast- and slow-quench limits and showed that, on the boundary, the coalescence of two critical modes gives rise to a tangential critical mode. While an ordinary crossing of $p_q=1/2$ produces the conventional linear cusp with critical exponent $\alpha=1$ and a jump in the dynamical winding number, the tangential mode produces a square-root singularity with $\alpha=1/2$ and a divergent first derivative of the rate function. Thus, the waiting time controls not only the number of critical modes but also the local critical behavior of the DQPT.

\begin{figure}[t]
  \begin{center}
		\includegraphics[width=3.3 in,angle=0]{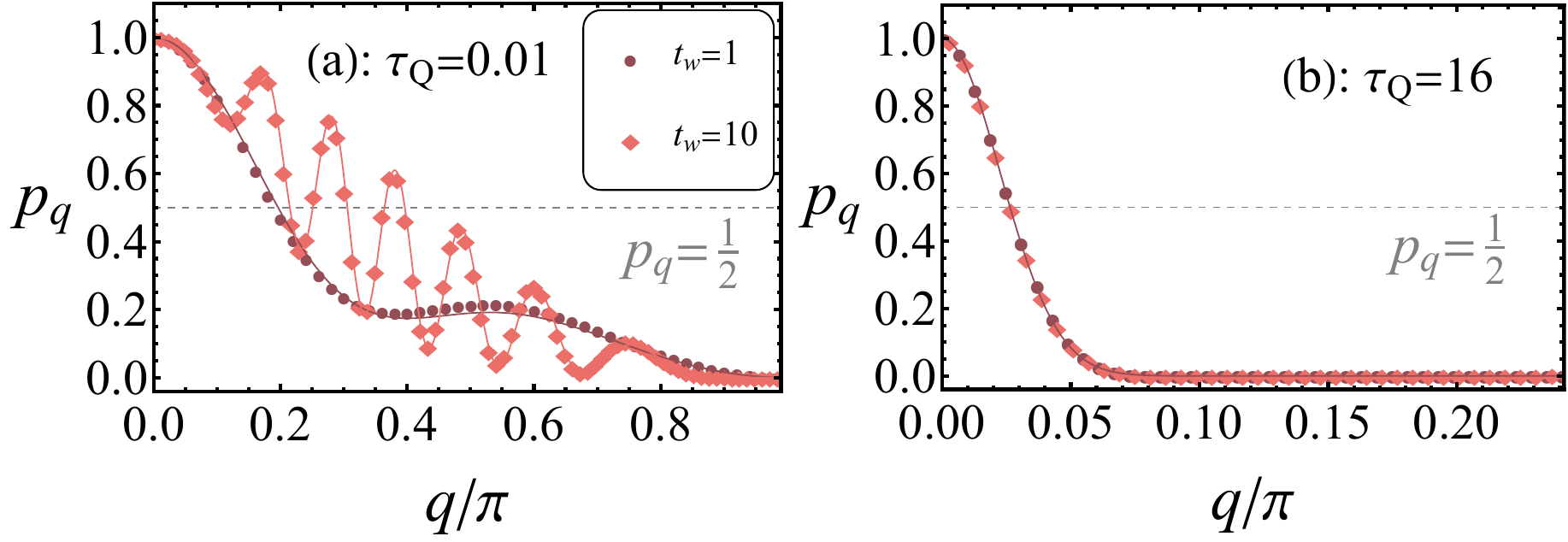}
  \end{center}
  \caption{Excitation probability for a waiting stage performed at $g_\text{wt}=0.5$ in the transverse Ising model. Panel (a) shows the fast-quench regime with $\tau_Q=0.01$. The excitation probability $p_q$ remains sensitive to the dynamical phase accumulated during the waiting stage and exhibits coherent oscillations. However, the oscillation amplitude is substantially reduced compared with the case in which the system waits at the critical point $g_c$. Panel (b) shows the slow-quench regime with $\tau_Q=16$. The excitation probability curves for different waiting times nearly overlap, indicating that the accumulated dynamical phase has a negligible effect on $p_q$ in this regime. The symbols denote the numerical results. The colored solid curves in panel (a) are obtained from Eq. \eqref{appeq-pqgwtfast}, whereas those in panel (b) are given by the Landau–Zener expression $p_q=e^{-2\pi\tau_Q q^2}$. The horizontal gray dashed line marks $p_q=1/2$.}
  \label{plot-pqgwt}
\end{figure}

The location of the waiting stage is crucial to this mechanism. When the system is held away from the critical point in the gapped ferromagnetic phase, the evolution remains coherent within each momentum sector; nevertheless, the finite excitation gap may weaken the coherent oscillations in the excitation probability \cite{Dziarmaga_2021, Dziarmaga_2022}. In App. \ref{App-pqgwt-waiting}, we derive the excitation probability for a waiting stage away from criticality under the quench protocol defined in Eq. \eqref{appeq-protocolgwt}, with the corresponding numerical results shown in Fig. \ref{plot-pqgwt}. Compared with waiting at the critical point $g_c$, the oscillation amplitude is substantially reduced for the fast quench, whereas the oscillations become essentially absent for the slow quench. Moreover, when the waiting stage is placed at the critical point, a sufficiently long waiting time can also reduce the final defect density, as shown in Fig. \ref{plot-DD}(a), indicating a reduced density of final excitations. This reduction originates from coherent interference between the two linear ramps rather than from a simple modification of KZ scaling. 
Future work should clarify the robustness of this interference mechanism against finite-size effects, weak interactions, integrability breaking, and environmental decoherence, and explore its extension to multiband and higher-dimensional systems.

\section*{ACKNOWLEDGMENTS}

We thank Siyan Lin for useful discussions. This work is supported by the Natural Science Foundation of Sichuan Province under Grant No. 2026NSFSC0772 (H.K.) and by the National Natural Science Foundation of China under Grant No. 11074177 (P.L.).

\section*{DATA AVAILABILITY}
The data that support the findings of this article are openly available \cite{Kou_data}.

\appendix

\section{Transverse Ising model and KZM} \label{App-sec-KZM}

After the Jordan-Wigner and Fourier transformations, the Hamiltonian, Eq. \eqref{H-Ising}, decomposes into independent momentum sectors, $H=\sum_q H_q$ with
\begin{equation}\label{eq-hqmode}
H_q= \epsilon_q (c_q^\dagger c_q-c_{-q}c_{-q}^\dagger)+\Delta_q (c_q^\dagger c_{-q}^\dagger+c_{-q}c_q),
\end{equation}
In Eq. \eqref{eq-hqmode}, $c_q$ and $c_q^\dagger$ are fermion annihilation and creation operators in momentum space, and 
\begin{align}
  \epsilon_q=2(g-\cos q), \Delta_q=2\sin q.
\end{align}

Using the Bogoliubov transformation $c_q=u_q\eta_q-v_q\eta_{-q}^\dagger$, the Hamiltonian is diagonalized as $H=\sum_q\omega_q(2\eta_q^\dagger\eta_q-1)$. $u_q$ and $v_q$ are Bogoliubov coefficients, and the quasiparticle dispersion reads $\omega_q=2\sqrt{1+g^2-2g\cos q}$. 

The final excitation probability is defined with respect to the instantaneous ground state of the post-quench Hamiltonian (at $t_f$),
\begin{equation}
 p_q=\langle \psi(t_f)|\eta_q^\dagger \eta_q|\psi(t_f)\rangle,
\label{eq-pq_def}
\end{equation}
where $|\psi(t_f)\rangle$ is the quantum state at $t_f$,
\begin{align}
  |\psi(t_f)\rangle=\prod_{q}[u_q(t_f)-v_q(t_f)c_q^\dagger c_{-q}^\dagger]|0\rangle,
\end{align}
and $|0\rangle$ is a fermion vacuum state, $c_q|0\rangle=0$. $u_q(t)$ and $v_q(t)$ are two time-dependent Bogoliubov coefficients, which are obtained from
\begin{align}\label{app-tdBdG}
i\frac{\mathrm{d}}{\mathrm{d}t}
\left[
\begin{array}{c}
u_{q}(t)\\v_{q}(t)
\end{array}
\right] =
\left[
\begin{array}{cc}
\epsilon_{q}(t)&\Delta_{q}\\\Delta_{q}&-\epsilon_{q}(t)
\end{array}
\right]
\left[
\begin{array}{c}
u_{q}(t)\\v_{q}(t)
\end{array}
\right],
\end{align}
The corresponding kink density is
\begin{equation}
 n=\frac{1}{\pi}\int_0^\pi p_q\,d q.
\label{eq-n}
\end{equation}

For $t_w=0$, the two-step linear quench protocol with an intermediate waiting stage reduces to a conventional one-step linear quench. In this case, the dynamical phase accumulated during the linear quench enters only as a common phase factor of the quantum state, and therefore has no effect on the excitation probability  \cite{Dziarmaga_2021}. For $g_f=0$, the excitation probability is given by
\begin{align}
  p_q&=p_q^\text{lz}=e^{-2\pi\tau_Q q^2}, ~~\tau_Q\gg1, \label{appeq-pqKZS}\\
  p_q&=p_q^\text{su}=\cos^2\frac{q}{2} -\frac{1}{2g_i}\sin^2 q,  ~~\tau_Q\ll1,  \label{appeq-pqsudden}
\end{align}
in the slow and fast quench limits, respectively. The corresponding defect densities are
\begin{align}
  n&=n_\text{KZ}=\frac{1}{2\pi\sqrt{2\tau_Q}},~~\tau_Q\gg1. \label{eq-nKZS} \\
  n&=n_\text{su}=\frac{1}{2}-\frac{1}{4g_i},~~\tau_Q\ll1, \label{eq-nsudd}
\end{align}
in the slow and fast limits, respectively.
As a consequence, the dynamics is described by the KZM for the slow quench \cite{Zurek_2005, Dziarmaga_2005, Polkovnikov_2005, Dziarmaga_2010}, whereas for the fast quench, it depends primarily on the initial field $g_i$ \cite{Campo_2023}. 

\section{Excitation probability after two-step linear quench with a waiting stage} \label{App-pq-waiting}

In Eq. \eqref{eq-protocol}, we adopt a two-step linear quench protocol with a waiting time $t_w$ at the critical point $g_c$. At $t_2$, the quantum state in the $q$ sector can be written as
\begin{align}
  |\psi_q(t_2)\rangle=\sqrt{1-A_q}e^{ia_q}e^{i\phi_q^\text{w}}|G_c\rangle+\sqrt{A_q}e^{i b_q}e^{-i\phi_q^\text{w}}|E_c\rangle,
\end{align}
where $|G_c\rangle$ and $|E_c\rangle$ are instantaneous ground and excited states at the critical point $g_c$, and $A_q$ is the excitation probability generated during the quench from $g_i$ to $g_c$. Both $a_q$ and $b_q$ are phase factors accumulated near the critical point during the linear quench, whereas $\phi_q^\text{w}$ is a dynamical phase accumulated during the waiting stage. We have
\begin{align}
  \phi_q^\text{w}=&-\int_{t_1}^{t_2}\langle G_c|H_q(g_c)|G_c\rangle dt'=\omega_q(g_c)t_w\nonumber\\
  =&4t_w\sin\frac{q}{2}.
\end{align}

After the second linear quench stage (from $t_2$ to $t_f$) is completed, the quantum state reads
\begin{align}\label{appeq-psitf}
  |\psi_q(t_f)\rangle=U_{cf}|\psi_q(t_2)\rangle,
\end{align}
where $U_{cf}$ is an evolution matrix for the one-way linear quench protocol from $g_c$ to $g_f$. Note that the basis vectors of state $|\psi_q(t_2)\rangle$ are $|G_c\rangle$ and $|E_c\rangle$. We have
\begin{align}
  U_{cf}|G_c\rangle&=\sqrt{1-B_q}e^{i c_q}|G_f\rangle +\sqrt{B_q}e^{i d_q}|E_f\rangle,\\
  U_{cf}|E_c\rangle&=-\sqrt{B_q}e^{-i d_q}|G_f\rangle +\sqrt{1-B_q}e^{-i c_q}|E_f\rangle,
\end{align} 
where $|G_f\rangle$ and $|E_f\rangle$ are instantaneous ground and excited states at the final field $g_f$, $B_q$ is the excitation probability for a one-way linear quench protocol from $g_c$ to $g_f$. Like $a_q$ and $b_q$, $c_q$ and $d_q$ are also phase factors accumulated near the critical point during the linear quench. Therefore, the evolution matrix reads
\begin{align} \label{appeq-ucf}
  U_{cf}=\left(
  \begin{array}{cc}
    \sqrt{1-B_q}e^{i c_q} & -\sqrt{B_q}e^{i d_q} \\
    \sqrt{B_q}e^{-i d_q} & \sqrt{1-B_q}e^{-i c_q} 
  \end{array}
  \right).
\end{align}
Substituting Eq. \eqref{appeq-ucf} into Eq. \eqref{appeq-psitf}, the final quantum state can be expressed in the basis spanned by $|G_f\rangle$ and $|E_f\rangle$, and the final excitation probability is defined as $p_q= |\langle E_f|\psi_q(t_f)\rangle|^2$, which reads 
\begin{align} \label{appeq-pqf}
  p_q= & (1-A_q)B_q+(1-B_q)A_q \nonumber\\
       & +2\sqrt{(1-A_q)B_q(1-B_q)A_q} \nonumber\\
       &\times \cos(2\phi_q^\text{w}-a_q-b_q-c_q-d_q).
\end{align}

First, for the fast quench limit $\tau_Q\rightarrow 0$, we have
\begin{align}
   A_q&=p_q^\text{su}|_{g_i\rightarrow g_c}\approx\frac{1}{2}\left(1-\sin\frac{q}{2}\right)-\frac{1}{g_i}\sin\frac{q}{2}\cos^2\frac{q}{2},\\
   B_q&=p_q^\text{su}|_{g_c\rightarrow 0}\approx\frac{1}{2}\left(1-\sin\frac{q}{2}\right),
\end{align}
and $a_q=b_q=c_q=d_q=0$. When $t_w=0$, the excitation probability, Eq. \eqref{appeq-pqf}, is that for a sudden quench from $g_i$ to $g_f$,
\begin{align}
  p_q=p_q^\text{su}|_{g_i\rightarrow 0}=\cos^2\frac{q}{2} -\frac{1}{2 g_i}\sin^2 q.
\end{align}
Therefore, in the limit of $g_i\gg 1$, the final excitation probability is shown in Eq. \eqref{eq-nsudd-waiting}.

Second, for the slow quench $\tau_Q\gg 1$, we have $A_q\approx B_q\approx p_q^\text{lz}(t_1)$ where $p_q^\text{lz}(t_1)$ is shown in Eq. \eqref{eq-pqlzt1} and denotes the excitation probability for a quench from the gapped phase to the critical point. In the long-wavelength limit, the dynamical phase accumulated near the critical point during the linear quench reads
\begin{align}
   a_q+b_q+c_q+d_q\approx - 2\sqrt{\pi \tau_Q}q.
\end{align}

\section{Critical exponent of DQPT} \label{app-criticalE}

In this section, we derive the local critical exponent of the rate function. Near $q^*$, we can expand the excitation probability with $p_{q}=\frac{1}{2}-a (q-q^*)$ or $p_{q}=\frac{1}{2}-a (q-q^*)^2$. We define $x=q-q^*$.

\subsection{$p_{q}=\frac{1}{2}-a (q-q^*)$}

The rate function reads 
\begin{align}
  r(\tilde{t})=&-\frac{1}{\pi}\int_{-\delta}^\delta dq\ln\left[1-\sin^2(2\tilde{t})+4a^2x^2 \sin^2(2\tilde{t})\right]\nonumber\\
  =&-\frac{1}{\pi}\left[2\delta\ln(A+B\delta^2)-4\delta+4\sqrt{\frac{A}{B}}\arctan\left(\delta\sqrt{\frac{B}{A}}\right)\right]
\end{align}
where $0<\delta\ll1$ sets the local momentum window, $A=\cos^2(2\tilde{t})$ and $B=4a^2\sin^2(2\tilde{t})$. At the critical time $\tilde{t}^*_n=\frac{(2n+1)\pi}{4}$ with $A=0$ and $B=4a^2$, the rate function shows nonanalytic behavior,
\begin{align}
  r(\tilde{t}^*_n)=-\frac{4\delta}{\pi}[\ln(2a\delta)-1].
\end{align}
Near $\tilde{t}^*_n$ with $A=4(\tilde{t}-\tilde{t}^*_n)^2$ and $B=4a^2$ ($A\ll B$), the rate function is
\begin{align}
  r(\tilde{t})=r(\tilde{t}^*_n)+4\frac{|\cos (2\tilde{t})|}{|2a \sin(2\tilde{t})| }\arctan\left(\delta\frac{|2a \sin(2\tilde{t})|}{|\cos (2\tilde{t})| }\right).
\end{align}
Because the term $\arctan\left(\delta\sqrt{\frac{B}{A}}\right)$ is a finite value, we have
\begin{align}
  r(\tilde{t})-r(\tilde{t}^*_n)\propto |\tilde{t}-\tilde{t}^*_n|.
\end{align}

\subsection{$p_{q}=\frac{1}{2}-a (q-q^\text{M})^2$}

The rate function reads 
\begin{align}
  r(\tilde{t})=&-\frac{1}{\pi}\int_{-\delta}^\delta dq\ln\left(A+Bx^4 \right)\nonumber\\
  =&-\frac{1}{\pi}\left[2\delta\ln(A+B\delta^4)-8\delta+8 b G(Y)\right]
\end{align}
where $A=\cos^2(2\tilde{t})$, $B=4a^2\sin^2(2\tilde{t})$, $b=(A/B)^{1/4}$, $Y=\delta/b$, and $G(Y)=\int_{0}^{Y}\frac{du}{1+u^4}$. At the critical time $\tilde{t}^*_n=\frac{(2n+1)\pi}{4}$, the rate function shows nonanalytic behavior,
\begin{align}
  r(\tilde{t}^*_n)=-\frac{4\delta}{\pi}[\ln(2a\delta^2)-2].
\end{align}
Near $\tilde{t}^*_n$ with $A=4(\tilde{t}-\tilde{t}^*)^2$ and $B=4a^2$ ($A\ll B$), the rate function is
\begin{align}
  r(\tilde{t})=r(\tilde{t}^*_n)-2\sqrt{2}\left(\frac{|\cos (2\tilde{t})|}{|2a \sin(2\tilde{t})| }\right)^{1/2}+O(\cos (2\tilde{t})).
\end{align}
We have
\begin{align}
  r(\tilde{t})-r(\tilde{t}^*_n)\propto |\tilde{t}-\tilde{t}^*_n|^{1/2}.
\end{align}

\section{Excitation probability for waiting stage away from criticality} \label{App-pqgwt-waiting}

In this section, we show the excitation probability in the transverse Ising chain when the waiting stage is placed away from the critical point,
\begin{align} \label{appeq-protocolgwt}
 g(t)=\left\{
 \begin{array}{cc}
 g_i-\frac{t}{\tau_Q},
 & 0\leq t<t_1,\\
 g_\text{wt},
 & t_1 \leq t\leq t_2,\\
 g_i-\frac{t-t_w}{\tau_Q},
 &  t_2 \leq t < t_f,
 \end{array}\right.
\end{align}
where $t_1=(g_i-g_\text{wt})\tau_Q$, $t_2=t_1+t_w$, $t_f= g_i\tau_Q+t_w$, and $g(t_f)=g_f=0$. $0<g_\text{wt}<g_c$ is the value of the transverse field during the waiting stage. In this case, the excitation probability can still be expressed in the form of Eq. \eqref{appeq-pqf}, but the accumulated dynamical phase reads
\begin{align}
  \phi_q^\text{w}&=-\int_{t_1}^{t_2}\langle G_\text{wt}| H_q(g_\text{wt})|G_\text{wt}\rangle dt'=\omega_q(g_\text{wt})t_w\nonumber\\
                 &=2\sqrt{1+g_\text{wt}^2-2g_\text{wt}\cos q}~t_w,
\end{align}
where $|G_\text{wt}\rangle$ is the ground state of $H_q(g_\text{wt})$.

In the fast quench limit $\tau_Q\rightarrow 0$, we have
\begin{align}
  A_q=&\frac{1}{2}\left[1-\frac{4(g_i-\cos q)(g_\text{wt}-\cos q)+4\sin^2q}{\omega_q(g_i)\omega_q(g_\text{wt})}\right],\\
  B_q=&\frac{1}{2}\left[1-\frac{-2\cos q(g_\text{wt}-\cos q)+2\sin^2q}{\omega_q(g_\text{wt})}\right].
\end{align}
The final excitation probability can be approximately expressed as
\begin{align}\label{appeq-pqgwtfast}
  p_q=&p_q^\text{su}|_{g_i\rightarrow 0}-4\sqrt{A_q(1-A_q)B_q(1-B_q)}\sin^2(\phi_q^\text{w})\nonumber\\
  =&\cos^2\frac{q}{2}-\frac{\sin^2q}{2g_i}-\frac{g_\text{wt}\sin^2q}{1+g_\text{wt}^2-2g_\text{wt}\cos q}\nonumber\\
  &\times \sqrt{1-\frac{2}{g_i}(g_\text{wt}-\cos q)}\sin^2\phi_q^\text{w}.
\end{align}

For the slow quench $\tau_Q\gg 1$, the first quench stage (from $g_i$ to $g_\text{wt}$) crosses the critical point, but the second quench stage does not cross the critical point. Therefore, we have $A_q=e^{-2\pi\tau_Q q^2}$ and $B_q=0$, and the final excitation probability $p_q=A_q=e^{-2\pi\tau_Q q^2}$.

\nocite{*}
\bibliographystyle{apsrev4-2}
\bibliography{citation}

@CONTROL{apsrev4-2,
  title = "1"
}

@article{Zurek_2005,
  title = {Dynamics of a Quantum Phase Transition},
  author = {Zurek, Wojciech H. and Dorner, Uwe and Zoller, Peter},
  journal = {Phys. Rev. Lett.},
  volume = {95},
  issue = {10},
  pages = {105701},
  numpages = {4},
  year = {2005},
  month = {Sep},
  publisher = {American Physical Society},
  doi = {10.1103/PhysRevLett.95.105701},
  url = {https://link.aps.org/doi/10.1103/PhysRevLett.95.105701}
}

@article{Dziarmaga_2005,
  title = {Dynamics of a Quantum Phase Transition: Exact Solution of the Quantum Ising Model},
  author = {Dziarmaga, Jacek},
  journal = {Phys. Rev. Lett.},
  volume = {95},
  issue = {24},
  pages = {245701},
  numpages = {4},
  year = {2005},
  month = {Dec},
  publisher = {American Physical Society},
  doi = {10.1103/PhysRevLett.95.245701},
  url = {https://link.aps.org/doi/10.1103/PhysRevLett.95.245701}
}

@article{Polkovnikov_2005,
  title = {Universal adiabatic dynamics in the vicinity of a quantum critical point},
  author = {Polkovnikov, Anatoli},
  journal = {Phys. Rev. B},
  volume = {72},
  issue = {16},
  pages = {161201 (R)},
  numpages = {4},
  year = {2005},
  month = {Oct},
  publisher = {American Physical Society},
  doi = {10.1103/PhysRevB.72.161201},
  url = {https://link.aps.org/doi/10.1103/PhysRevB.72.161201}
}

@article{Sen_2010,
  title = {Landau-Zener problem with waiting at the minimum gap and related quench dynamics of a many-body system},
  author = {Divakaran, Uma and Dutta, Amit and Sen, Diptiman},
  journal = {Phys. Rev. B},
  volume = {81},
  issue = {5},
  pages = {054306},
  numpages = {6},
  year = {2010},
  month = {Feb},
  publisher = {American Physical Society},
  doi = {10.1103/PhysRevB.81.054306},
  url = {https://link.aps.org/doi/10.1103/PhysRevB.81.054306}
}

@article{Dziarmaga_2010,
  title = {Dynamics of a quantum phase transition and relaxation to a steady state},
  author = {Dziarmaga, Jacek},
 journal = {Advances in Physics},
  volume = {59},
  issue = {6},
  pages = {1063-1189},
  numpages = {6},
  year = {2010},
  month = {November},
  publisher = {Taylor \& Francis},
  doi = {10.1080/00018732.2010.514702},
  url = {https://doi.org/10.1080/00018732.2010.514702}
}

@article{Nori_2010,
  title = {Landau–Zener–Stückelberg interferometry},
  author = {S.N. Shevchenko and S. Ashhab and Franco Nori},
  journal = {Physics Reports},
  volume = {492},
  issue = {},
  pages = {1-30},
  numpages = {1},
  year = {2010},
  month = {},
  publisher = {},
  doi = {https://doi.org/10.1016/j.physrep.2010.03.002},
  url = {https://www.sciencedirect.com/science/article/pii/S0370157310000815}
}

@article{Hely_2013,
  title = {Dynamical Quantum Phase Transitions in the Transverse-Field Ising Model},
  author = {Heyl, M. and Polkovnikov, A. and Kehrein, S.},
  journal = {Phys. Rev. Lett.},
  volume = {110},
  issue = {13},
  pages = {135704},
  numpages = {5},
  year = {2013},
  month = {Mar},
  publisher = {American Physical Society},
  doi = {10.1103/PhysRevLett.110.135704},
  url = {https://link.aps.org/doi/10.1103/PhysRevLett.110.135704}
}

@article{Hely_2015PRL,
  title = {Scaling and Universality at Dynamical Quantum Phase Transitions},
  author = {Heyl, Markus},
  journal = {Phys. Rev. Lett.},
  volume = {115},
  issue = {14},
  pages = {140602},
  numpages = {5},
  year = {2015},
  month = {Oct},
  publisher = {American Physical Society},
  doi = {10.1103/PhysRevLett.115.140602},
  url = {https://link.aps.org/doi/10.1103/PhysRevLett.115.140602}
}

@article{Markus_2015,
  title = {Dynamical quantum phase transitions in the Kitaev honeycomb model},
  author = {Schmitt, Markus and Kehrein, Stefan},
  journal = {Phys. Rev. B},
  volume = {92},
  issue = {7},
  pages = {075114},
  numpages = {13},
  year = {2015},
  month = {Aug},
  publisher = {American Physical Society},
  doi = {10.1103/PhysRevB.92.075114},
  url = {https://link.aps.org/doi/10.1103/PhysRevB.92.075114}
}

@article{Dirk_2016,
	title = {Time evolution during and after finite-time quantum quenches in the transverse-field Ising chain},
	pages = {003},
	author = {Puskarov, Tatjana and Schuricht, Dirk},
	journal = {SciPost Phys.},
	volume = {1},
	year = {2016},
	publisher = {SciPost},
	doi = {10.21468/SciPostPhys.1.1.003},
	url = {https://scipost.org/10.21468/SciPostPhys.1.1.003}
}

@article{Hely_2016_DTOP,
  title = {Dynamical topological order parameters far from equilibrium},
  author = {Budich, Jan Carl and Heyl, Markus},
  journal = {Phys. Rev. B},
  volume = {93},
  issue = {8},
  pages = {085416},
  numpages = {7},
  year = {2016},
  month = {Feb},
  publisher = {American Physical Society},
  doi = {10.1103/PhysRevB.93.085416},
  url = {https://link.aps.org/doi/10.1103/PhysRevB.93.085416}
}

@article{Polkovnikov_2016,
  title = {Slow quenches in a quantum Ising chain: Dynamical phase transitions and topology},
  author = {Sharma, Shraddha and Divakaran, Uma and Polkovnikov, Anatoli and Dutta, Amit},
  journal = {Phys. Rev. B},
  volume = {93},
  issue = {14},
  pages = {144306},
  numpages = {9},
  year = {2016},
  month = {Apr},
  publisher = {American Physical Society},
  doi = {10.1103/PhysRevB.93.144306},
  url = {https://link.aps.org/doi/10.1103/PhysRevB.93.144306}
}

@article{Huang_2016,
  title = {Dynamical Quantum Phase Transitions: Role of Topological Nodes in Wave Function Overlaps},
  author = {Huang, Zhoushen and Balatsky, Alexander V.},
  journal = {Phys. Rev. Lett.},
  volume = {117},
  issue = {8},
  pages = {086802},
  numpages = {6},
  year = {2016},
  month = {Aug},
  publisher = {American Physical Society},
  doi = {10.1103/PhysRevLett.117.086802},
  url = {https://link.aps.org/doi/10.1103/PhysRevLett.117.086802}
}

@article{Hely_2017,
  title = {Direct Observation of Dynamical Quantum Phase Transitions in an Interacting Many-Body System},
  author = {Jurcevic, P. and Shen, H. and Hauke, P. and Maier, C. and Brydges, T. and Hempel, C. and Lanyon, B. P. and Heyl, M. and Blatt, R. and Roos, C. F.},
  journal = {Phys. Rev. Lett.},
  volume = {119},
  issue = {8},
  pages = {080501},
  numpages = {5},
  year = {2017},
  month = {Aug},
  publisher = {American Physical Society},
  doi = {10.1103/PhysRevLett.119.080501},
  url = {https://link.aps.org/doi/10.1103/PhysRevLett.119.080501}
}

@article{Hely_2018,
  title = {Observation of dynamical vortices after quenches in a system with topology},
  author = {Fläschner, N. and Vogel, D. and Tarnowski, M. and Rem, B. S. and Lühmann, D.-S. and Heyl, M. and Budich, J. C. and Mathey, L. and Sengstock, K. and Weitenberg, C.},
  journal = {Nature Physics},
  volume = {14},
  issue = {3},
  pages = {265-268},
  numpages = {},
  year = {2018},
  month = {Mar},
  publisher = {},
  doi = {10.1038/s41567-017-0013-8},
  url = {https://doi.org/10.1038/s41567-017-0013-8}
}

@article{Dutta_2018,
  title = {Dynamical quantum phase transitions in extended transverse Ising models},
  author = {Bhattacharjee, Sourav and Dutta, Amit},
  journal = {Phys. Rev. B},
  volume = {97},
  issue = {13},
  pages = {134306},
  numpages = {8},
  year = {2018},
  month = {Apr},
  publisher = {American Physical Society},
  doi = {10.1103/PhysRevB.97.134306},
  url = {https://link.aps.org/doi/10.1103/PhysRevB.97.134306}
}

@article{Heyl_2018,
  title = {Dynamical quantum phase transitions: a review},
  author = {Heyl, Markus},
  journal = {Reports on Progress in Physics},
  volume = {81},
  issue = {},
  pages = {054001},
  numpages = {5},
  year = {2018},
  month = {apr},
  publisher = {IOP Publishing},
  doi = {10.1088/1361-6633/aaaf9a},
  url = {https://doi.org/10.1088/1361-6633/aaaf9a}
}

@article{Kennes_2018,
  title = {Controlling dynamical quantum phase transitions},
  author = {Kennes, D. M. and Schuricht, D. and Karrasch, C.},
  journal = {Phys. Rev. B},
  volume = {97},
  issue = {18},
  pages = {184302},
  numpages = {6},
  year = {2018},
  month = {May},
  publisher = {American Physical Society},
  doi = {10.1103/PhysRevB.97.184302},
  url = {https://link.aps.org/doi/10.1103/PhysRevB.97.184302}
}

@article{Heyl_2018-2,
  title = {Detecting Equilibrium and Dynamical Quantum Phase Transitions in Ising Chains via Out-of-Time-Ordered Correlators},
  author = {Heyl, Markus and Pollmann, Frank and D\'ora, Bal\'azs},
  journal = {Phys. Rev. Lett.},
  volume = {121},
  issue = {1},
  pages = {016801},
  numpages = {7},
  year = {2018},
  month = {Jul},
  publisher = {American Physical Society},
  doi = {10.1103/PhysRevLett.121.016801},
  url = {https://link.aps.org/doi/10.1103/PhysRevLett.121.016801}
}

@article{XueP_2019,
  title = {Simulating Dynamic Quantum Phase Transitions in Photonic Quantum Walks},
  author = {Wang, Kunkun and Qiu, Xingze and Xiao, Lei and Zhan, Xiang and Bian, Zhihao and Yi, Wei and Xue, Peng},
  journal = {Phys. Rev. Lett.},
  volume = {122},
  issue = {2},
  pages = {020501},
  numpages = {6},
  year = {2019},
  month = {Jan},
  publisher = {American Physical Society},
  doi = {10.1103/PhysRevLett.122.020501},
  url = {https://link.aps.org/doi/10.1103/PhysRevLett.122.020501}
}

@article{ChenS_2019,
  title = {Observation of a Dynamical Quantum Phase Transition by a Superconducting Qubit Simulation},
  author = {Guo, Xue-Yi and Yang, Chao and Zeng, Yu and Peng, Yi and Li, He-Kang and Deng, Hui and Jin, Yi-Rong and Chen, Shu and Zheng, Dongning and Fan, Heng},
  journal = {Phys. Rev. Appl.},
  volume = {11},
  issue = {4},
  pages = {044080},
  numpages = {12},
  year = {2019},
  month = {Apr},
  publisher = {American Physical Society},
  doi = {10.1103/PhysRevApplied.11.044080},
  url = {https://link.aps.org/doi/10.1103/PhysRevApplied.11.044080}
}

@article{DuanLM_2020,
  title = {Observation of Dynamical Quantum Phase Transitions with Correspondence in an Excited State Phase Diagram},
  author = {Tian, T. and Yang, H.-X. and Qiu, L.-Y. and Liang, H.-Y. and Yang, Y.-B. and Xu, Y. and Duan, L.-M.},
  journal = {Phys. Rev. Lett.},
  volume = {124},
  issue = {4},
  pages = {043001},
  numpages = {6},
  year = {2020},
  month = {Jan},
  publisher = {American Physical Society},
  doi = {10.1103/PhysRevLett.124.043001},
  url = {https://link.aps.org/doi/10.1103/PhysRevLett.124.043001}
}

@article{WuYT_2020,
  title = {Dynamical quantum phase transitions of quantum spin chains with a Loschmidt-rate critical exponent equal to $\frac{1}{2}$},
  author = {Wu, Yantao},
  journal = {Phys. Rev. B},
  volume = {101},
  issue = {6},
  pages = {064427},
  numpages = {6},
  year = {2020},
  month = {Feb},
  publisher = {American Physical Society},
  doi = {10.1103/PhysRevB.101.064427},
  url = {https://link.aps.org/doi/10.1103/PhysRevB.101.064427}
}

@article{Porta_2020,
  title = {Topological classification of dynamical quantum phase transitions in the xy chain},
  author = {Porta, Sergio and Cavaliere, Fabio and Sassetti, Maura and Traverso Ziani, Niccolò},
  journal = {Scientific Reports},
  volume = {10},
  issue = {1},
  pages = {12766},
  numpages = {},
  year = {2020},
  month = {July},
  publisher = {},
  doi = {10.1038/s41598-020-69621-8},
  url = {https://doi.org/10.1038/s41598-020-69621-8}
}

@article{Ding_2020,
  title = {Dynamical quantum phase transition from a critical quantum quench},
  author = {Ding, Chengxiang},
  journal = {Phys. Rev. B},
  volume = {102},
  issue = {6},
  pages = {060409(R)},
  numpages = {5},
  year = {2020},
  month = {Aug},
  publisher = {American Physical Society},
  doi = {10.1103/PhysRevB.102.060409},
  url = {https://link.aps.org/doi/10.1103/PhysRevB.102.060409}
}

@article{Link_2020,
  title = {Dynamical Phase Transitions in Dissipative Quantum Dynamics with Quantum Optical Realization},
  author = {Link, Valentin and Strunz, Walter T.},
  journal = {Phys. Rev. Lett.},
  volume = {125},
  issue = {14},
  pages = {143602},
  numpages = {6},
  year = {2020},
  month = {Sep},
  publisher = {American Physical Society},
  doi = {10.1103/PhysRevLett.125.143602},
  url = {https://link.aps.org/doi/10.1103/PhysRevLett.125.143602}
}

@article{GuoGC_2020,
  title = {Measuring a dynamical topological order parameter in quantum walks},
  author = {Xu, Xiao-Ye and Wang, Qin-Qin and Heyl, Markus and Budich, Jan Carl and Pan, Wei-Wei and Chen, Zhe and Jan, Munsif and Sun, Kai and Xu, Jin-Shi and Han, Yong-Jian and Li, Chuan-Feng and Guo, Guang-Can},
  journal = { Light: Science \& Applications},
  volume = {9},
  issue = {1},
  pages = {7},
  numpages = {},
  year = {2020},
  month = {Jan},
  publisher = {},
  doi = {10.1038/s41377-019-0237-8},
  url = {https://doi.org/10.1038/s41377-019-0237-8}
}

@article{XuK_2020,
  title = {Probing dynamical phase transitions with a superconducting quantum simulator},
  author = {Kai Xu  and Zheng-Hang Sun  and Wuxin Liu  and Yu-Ran Zhang  and Hekang Li  and Hang Dong  and Wenhui Ren  and Pengfei Zhang  and Franco Nori  and Dongning Zheng  and Heng Fan  and H. Wang},
  journal = { Science Advances},
  volume = {6},
  issue = {},
  pages = {eaba4935},
  numpages = {25},
  year = {2020},
  month = {},
  publisher = {},
  doi = {10.1126/sciadv.aba4935},
  url = {https://www.science.org/doi/abs/10.1126/sciadv.aba4935}
}

@article{Dziarmaga_2021,
  title = {Quantum Kibble-Zurek mechanism: Kink correlations after a quench in the quantum Ising chain},
  author = {Nowak, Rados\l{}aw J. and Dziarmaga, Jacek},
  journal = {Phys. Rev. B},
  volume = {104},
  issue = {7},
  pages = {075448},
  numpages = {10},
  year = {2021},
  month = {Aug},
  publisher = {American Physical Society},
  doi = {10.1103/PhysRevB.104.075448},
  url = {https://link.aps.org/doi/10.1103/PhysRevB.104.075448}
}

@article{Polkovnikov_2021,
  title = {Observing Dynamical Quantum Phase Transitions through Quasilocal String Operators},
  author = {Bandyopadhyay, Souvik and Polkovnikov, Anatoli and Dutta, Amit},
  journal = {Phys. Rev. Lett.},
  volume = {126},
  issue = {20},
  pages = {200602},
  numpages = {6},
  year = {2021},
  month = {May},
  publisher = {American Physical Society},
  doi = {10.1103/PhysRevLett.126.200602},
  url = {https://link.aps.org/doi/10.1103/PhysRevLett.126.200602}
}

@article{Peotta_2021PRX,
  title = {Determination of Dynamical Quantum Phase Transitions in Strongly Correlated Many-Body Systems Using Loschmidt Cumulants},
  author = {Peotta, Sebastiano and Brange, Fredrik and Deger, Aydin and Ojanen, Teemu and Flindt, Christian},
  journal = {Phys. Rev. X},
  volume = {11},
  issue = {4},
  pages = {041018},
  numpages = {15},
  year = {2021},
  month = {Oct},
  publisher = {American Physical Society},
  doi = {10.1103/PhysRevX.11.041018},
  url = {https://link.aps.org/doi/10.1103/PhysRevX.11.041018}
}

@article{Dziarmaga_2022,
  title = {Kink correlations, domain-size distribution, and emptiness formation probability after a Kibble-Zurek quench in the quantum Ising chain},
  author = {Dziarmaga, Jacek and Rams, Marek M.},
  journal = {Phys. Rev. B},
  volume = {106},
  issue = {1},
  pages = {014309},
  numpages = {14},
  year = {2022},
  month = {Jul},
  publisher = {American Physical Society},
  doi = {10.1103/PhysRevB.106.014309},
  url = {https://link.aps.org/doi/10.1103/PhysRevB.106.014309}
}

@article{kou_2022,
  title = {Interferometry based on the quantum Kibble-Zurek mechanism},
  author = {Kou, Han-Chuan and Li, Peng},
  journal = {Phys. Rev. B},
  volume = {106},
  issue = {18},
  pages = {184301},
  numpages = {18},
  year = {2022},
  month = {Nov},
  publisher = {American Physical Society},
  doi = {10.1103/PhysRevB.106.184301},
  url = {https://link.aps.org/doi/10.1103/PhysRevB.106.184301}
}

@article{Chien_2022,
  title = {Metamorphic dynamical quantum phase transition in double-quench processes at finite temperatures},
  author = {Hou, Xu-Yang and Gao, Qu-Cheng and Guo, Hao and Chien, Chih-Chun},
  journal = {Phys. Rev. B},
  volume = {106},
  issue = {1},
  pages = {014301},
  numpages = {9},
  year = {2022},
  month = {Jul},
  publisher = {American Physical Society},
  doi = {10.1103/PhysRevB.106.014301},
  url = {https://link.aps.org/doi/10.1103/PhysRevB.106.014301}
}

@article{Green_2022,
  title = {Simulating groundstate and dynamical quantum phase transitions on a superconducting quantum computer},
  author = {Dborin, James and Wimalaweera, Vinul and Barratt, F. and Ostby, Eric and O’Brien, Thomas E. and Green, A. G.},
  journal = {Nature Communications},
  volume = {13},
  issue = {1},
  pages = {5977},
  numpages = {},
  year = {2022},
  month = {Oct},
  publisher = {},
  doi = {10.1038/s41467-022-33737-4},
  url = {https://doi.org/10.1038/s41467-022-33737-4}
}

@article{Campo_2023,
  title = {Universal Breakdown of Kibble-Zurek Scaling in Fast Quenches across a Phase Transition},
  author = {Zeng, Hua-Bi and Xia, Chuan-Yin and del Campo, Adolfo},
  journal = {Phys. Rev. Lett.},
  volume = {130},
  issue = {6},
  pages = {060402},
  numpages = {6},
  year = {2023},
  month = {Feb},
  publisher = {American Physical Society},
  doi = {10.1103/PhysRevLett.130.060402},
  url = {https://link.aps.org/doi/10.1103/PhysRevLett.130.060402}
}

@article{Cao_2023,
  title = {Dynamics of the geometric phase in inhomogeneous quantum spin chains},
  author = {Cao, Kaiyuan and Yang, Shuxiang and Hu, Yayun and Yang, Guangwen},
  journal = {Phys. Rev. B},
  volume = {108},
  issue = {2},
  pages = {024201},
  numpages = {13},
  year = {2023},
  month = {Jul},
  publisher = {American Physical Society},
  doi = {10.1103/PhysRevB.108.024201},
  url = {https://link.aps.org/doi/10.1103/PhysRevB.108.024201}
}

@article{Cheraghi_2023,
  title = {Dynamical quantum phase transitions following double quenches: persistence of the initial state vs dynamical phases},
  author = {Cheraghi, Hadi and Sedlmayr, Nicholas},
  journal = {New Journal of Physics},
  volume = {25},
  issue = {},
  pages = {103035},
  numpages = {10},
  year = {2023},
  month = {oct},
  publisher = {IOP Publishing},
  doi = {10.1088/1367-2630/ad016e},
  url = {https://doi.org/10.1088/1367-2630/ad016e}
}

@article{Nori_2023,
  title = {Nonadiabatic Landau–Zener–Stückelberg–Majorana transitions, dynamics, and interference},
  author = {Oleh V. Ivakhnenko and Sergey N. Shevchenko and Franco Nori},
  journal = {Physics Reports},
  volume = {995},
  issue = {},
  pages = {1-89},
  numpages = {},
  year = {2023},
  month = {},
  publisher = {},
  doi = {https://doi.org/10.1016/j.physrep.2022.10.002},
  url = {https://www.sciencedirect.com/science/article/pii/S0370157322003702}
}

@article{LiP_2023,
  title = {Varying quench dynamics in the transverse Ising chain: The Kibble-Zurek, saturated, and presaturated regimes},
  author = {Kou, Han-Chuan and Li, Peng},
  journal = {Phys. Rev. B},
  volume = {108},
  issue = {21},
  pages = {214307},
  numpages = {11},
  year = {2023},
  month = {Dec},
  publisher = {American Physical Society},
  doi = {10.1103/PhysRevB.108.214307},
  url = {https://link.aps.org/doi/10.1103/PhysRevB.108.214307}
}

@article{Ding_2024,
  title = {Quench dynamics of an open quantum system},
  author = {Ding, Chengxiang and Zhang, Long},
  journal = {Phys. Rev. B},
  volume = {111},
  issue = {18},
  pages = {184319},
  numpages = {14},
  year = {2025},
  month = {May},
  publisher = {American Physical Society},
  doi = {10.1103/PhysRevB.111.184319},
  url = {https://link.aps.org/doi/10.1103/PhysRevB.111.184319}
}

@article{Chi_2024-apr,
  title = {Distribution of Fisher zeros in dynamical quantum phase transitions of two-dimensional topological systems},
  author = {Sacramento, P. D. and Yu, Wing Chi},
  journal = {Phys. Rev. B},
  volume = {109},
  issue = {13},
  pages = {134301},
  numpages = {18},
  year = {2024},
  month = {Apr},
  publisher = {American Physical Society},
  doi = {10.1103/PhysRevB.109.134301},
  url = {https://link.aps.org/doi/10.1103/PhysRevB.109.134301}
}

@article{Jafari_2024Aug,
  title = {Competition of long-range interactions and noise at a ramped quench dynamical quantum phase transition: The case of the long-range pairing Kitaev chain},
  author = {Baghran, R. and Jafari, R. and Langari, A.},
  journal = {Phys. Rev. B},
  volume = {110},
  issue = {6},
  pages = {064302},
  numpages = {11},
  year = {2024},
  month = {Aug},
  publisher = {American Physical Society},
  doi = {10.1103/PhysRevB.110.064302},
  url = {https://link.aps.org/doi/10.1103/PhysRevB.110.064302}
}

@article{Jafari_2024May,
  title = {Dynamical quantum phase transitions following a noisy quench},
  author = {Jafari, R. and Langari, A. and Eggert, S. and Johannesson, Henrik},
  journal = {Phys. Rev. B},
  volume = {109},
  issue = {18},
  pages = {L180303},
  numpages = {7},
  year = {2024},
  month = {May},
  publisher = {American Physical Society},
  doi = {10.1103/PhysRevB.109.L180303},
  url = {https://link.aps.org/doi/10.1103/PhysRevB.109.L180303}
}

@article{Chi_2024,
  title = {Entanglement in quenched extended Su-Schrieffer-Heeger model with anomalous dynamical quantum phase transitions},
  author = {Wong, Cheuk Yiu and Hui, Tsz Hin and Sacramento, P. D. and Yu, Wing Chi},
  journal = {Phys. Rev. B},
  volume = {110},
  issue = {5},
  pages = {054312},
  numpages = {17},
  year = {2024},
  month = {Aug},
  publisher = {American Physical Society},
  doi = {10.1103/PhysRevB.110.054312},
  url = {https://link.aps.org/doi/10.1103/PhysRevB.110.054312}
}

@article{CaoKY_2024,
  title = {Exploring dynamical phase transitions in the $XY$ chain through a linear quench: Early and long-term perspectives},
  author = {Cao, Kaiyuan and Hou, Hongsheng and Tong, Peiqing},
  journal = {Phys. Rev. A},
  volume = {110},
  issue = {4},
  pages = {042209},
  numpages = {12},
  year = {2024},
  month = {Oct},
  publisher = {American Physical Society},
  doi = {10.1103/PhysRevA.110.042209},
  url = {https://link.aps.org/doi/10.1103/PhysRevA.110.042209}
}

@article{ZhengZY_2025,
  title = {Loschmidt echo zeros in finite-size quantum systems with linear quench},
  author = {Zheng, Zhen-Yu and Liu, Xudong and Lin, Siyan and Zhang, Yu and Chen, Shu},
  journal = {Phys. Rev. B},
  volume = {113},
  issue = {2},
  pages = {L020301},
  numpages = {7},
  year = {2026},
  month = {Jan},
  publisher = {American Physical Society},
  doi = {10.1103/qc3w-s38g},
  url = {https://link.aps.org/doi/10.1103/qc3w-s38g}
}

@misc{Jafari_2025,
      title={Noise-Affected Dynamical Quantum Phase Transitions}, 
      author={R. Jafari and Alireza Akbari and Mehdi Biderang and Jesko Sirker},
      year={2025},
      eprint={2504.03005},
      archivePrefix={arXiv},
      primaryClass={cond-mat.stat-mech},
      url={https://arxiv.org/abs/2504.03005}, 
}

@article{kou_2025,
  title = {Kibble-Zurek scaling immune to anti-Kibble-Zurek behavior in driven open systems at the limit of loss difference},
  author = {Kou, Han-Chuan and Zhang, Zhi-Han and Li, Peng},
  journal = {Phys. Rev. B},
  volume = {111},
  issue = {15},
  pages = {155152},
  numpages = {14},
  year = {2025},
  month = {Apr},
  publisher = {American Physical Society},
  doi = {10.1103/PhysRevB.111.155152},
  url = {https://link.aps.org/doi/10.1103/PhysRevB.111.155152}
}

@article{Kiss_2025,
doi = {10.1088/2058-9565/addf75},
url = {https://doi.org/10.1088/2058-9565/addf75},
year = {2025},
month = {jun},
publisher = {IOP Publishing},
volume = {10},
number = {3},
pages = {035037},
author = {Kiss, Oriel and Teplitskiy, Daniil and Grossi, Michele and Mandarino, Antonio},
title = {Statistics of topological defects across a phase transition in a digital superconducting quantum processor},
journal = {Quantum Science and Technology}
}

@article{LiP_2025,
  title = {Critical dynamics and its interferometry in the one-dimensional $p$-wave-paired Aubry-Andr\'e-Harper model},
  author = {Zhang, Zhi-Han and Kou, Han-Chuan and Li, Peng},
  journal = {Phys. Rev. B},
  volume = {112},
  issue = {1},
  pages = {014310},
  numpages = {11},
  year = {2025},
  month = {Jul},
  publisher = {American Physical Society},
  doi = {10.1103/6331-k7vl},
  url = {https://link.aps.org/doi/10.1103/6331-k7vl}
}

@article{LiFX_2025,
  title = {Universal scaling in Loschmidt echo across quantum phase transitions under nonadiabatic dynamics},
  author = {Zhang, Xiang and Hu, Liangdong and Li, Fuxiang},
  journal = {Phys. Rev. B},
  volume = {112},
  issue = {2},
  pages = {024310},
  numpages = {14},
  year = {2025},
  month = {Jul},
  publisher = {American Physical Society},
  doi = {10.1103/lq1s-wx4d},
  url = {https://link.aps.org/doi/10.1103/lq1s-wx4d}
}

@article{XueP_2025,
  title = {Self-normal and biorthogonal dynamical quantum phase transitions in non-Hermitian quantum walks},
  author = {Zhang, Haiting and Wang, Kunkun and Xiao, Lei and Xue, Peng},
  journal = {Light: Science \& Applications},
  volume = {14},
  issue = {1},
  pages = {253},
  numpages = {},
  year = {2025},
  month = {July},
  publisher = {},
  doi = {10.1038/s41377-025-01919-6},
  url = {https://doi.org/10.1038/s41377-025-01919-6}
}

@article{Kou_2026,
  title = {Optimized adiabatic-impulse protocol preserving Kibble-Zurek scaling with attenuated anti-Kibble-Zurek behavior},
  author = {Kou, Han-Chuan and Zhang, Zhi-Han and Wu, Xin-Hui and Zhou, Yan and Chen, Gang and Li, Peng},
  journal = {Phys. Rev. B},
  volume = {113},
  issue = {17},
  pages = {174308},
  numpages = {12},
  year = {2026},
  month = {May},
  publisher = {American Physical Society},
  doi = {10.1103/5xrs-z9ls},
  url = {https://link.aps.org/doi/10.1103/5xrs-z9ls}
}

@misc{Cao_2026,
      title={Maximal Entanglement and Frozen Information: A Unified Framework for Dynamical Quantum Phase Transitions}, 
      author={Kaiyuan Cao and Mingzhi Li and Xiang-Ping Jiang and Shu Chen and Jian Wang},
      year={2026},
      eprint={2601.04535},
      archivePrefix={arXiv},
      primaryClass={quant-ph},
      url={https://arxiv.org/abs/2601.04535}, 
}

@misc{Bian_2026,
      title={Temporal Dynamical Quantum Phase Transition in Dicke Model with Trapped Ions}, 
      author={Ji Bian and Wei Wu and Zihan Xie and Mengxiang Zhang and Yi Li and Yue Li and Rixin Yao and Yuqi Zhou and Xu Cheng and Han Pu and Yiheng Lin},
      year={2026},
      eprint={2606.28664},
      archivePrefix={arXiv},
      primaryClass={quant-ph},
      url={https://arxiv.org/abs/2606.28664}, 
}

@article{Elisa_2026,
  title = {Digital Rydberg simulation of dynamical quantum phase transitions in the Schwinger model},
  author = {Pomarico, Domenico and Dell'Anna, Federico and Cioli, Riccardo and Pascazio, Saverio and Pepe, Francesco V. and Facchi, Paolo and Ercolessi, Elisa},
  journal = {Phys. Rev. Res.},
  volume = {8},
  issue = {1},
  pages = {013177},
  numpages = {13},
  year = {2026},
  month = {Feb},
  publisher = {American Physical Society},
  doi = {10.1103/b233-qtvm},
  url = {https://link.aps.org/doi/10.1103/b233-qtvm}
}

@article{YinS_2026,
  title = {Finite-time scaling with two characteristic time scales: Driven critical dynamics with emergent symmetry},
  author = {Shu, Yu-Rong and Yang, Li-Ying and Yin, Shuai},
  journal = {Phys. Rev. B},
  volume = {113},
  issue = {13},
  pages = {134303},
  numpages = {14},
  year = {2026},
  month = {Apr},
  publisher = {American Physical Society},
  doi = {10.1103/scjr-r3hj},
  url = {https://link.aps.org/doi/10.1103/scjr-r3hj}
}

@article{Kou_data,
  title = {{Dataset, Zenodo} (2026)},
  author = {Kou, Han-Chuan},
  journal = {doi: 10.5281/zenodo.23257129},
  volume = {},
  issue = {},
  pages = {},
  numpages = {},
  year = {},
  month = {},
  publisher = {Zenodo},
  doi = {10.5281/zenodo.23257129},
  url = {https://doi.org/10.5281/zenodo.23257129}
}

\end{document}